\documentclass[apj]{emulateapj}       
\usepackage{apjfonts,epsfig,natbib}   

\newcommand{\hi}{\hbox{\ion{H}{1}}}         
\newcommand{\hii}{\hbox{\ion{H}{2}}}        
\newcommand{\lz}{\hbox{$L$--$Z$}}
\newcommand{\mz}{\hbox{$M_{\ast}$--$Z$}}
\newcommand{\mstar}{\hbox{$M_{\ast}$}}
\newcommand{\vk}{\hbox{$V\!-\!K$}}
\newcommand{\bk}{\hbox{$B\!-\!K$}}
\newcommand{\btwo}{\hbox{$B\!-\![4.5]$}}
\newcommand{\ktwo}{\hbox{$K\!-\![4.5]$}}

\newcommand{\msun}{\hbox{M$_{\odot}$}}

\newcommand{\zsun}{\hbox{$Z_{\odot}$}}

\newcommand{\othreea}{\hbox{[\ion{O}{3}] $\lambda4363$}}

\newcommand{\ntwot}{\hbox{[\ion{N}{2}] $\lambda5755$}}

\journalinfo{ApJ, accepted (1 May 2006)}        

\shortauthors{H. Lee et al.}
\shorttitle{Stellar Mass-Metallicity Relation for Dwarf Irregulars}
\slugcomment{
Full version available at 
http://www.astro.umn.edu/$\sim$hlee/papers/
}

\begin{document}

\title{
On Extending the Mass-Metallicity Relation 
of Galaxies by 2.5 Decades in Stellar Mass
}

\author{
Henry Lee$\,$\altaffilmark{1},
Evan D. Skillman$\,$\altaffilmark{1},
John M. Cannon$\,$\altaffilmark{2},  
Dale C. Jackson$\,$\altaffilmark{1}, \\
Robert D. Gehrz$\,$\altaffilmark{1},
Elisha F. Polomski$\,$\altaffilmark{1},
and
Charles E. Woodward$\,$\altaffilmark{1}
}

\altaffiltext{1}{
Department of Astronomy, University of Minnesota,
116 Church St. S.E., Minneapolis, MN 55455;
hlee, skillman, djackson, gehrz, elwood, chelsea@astro.umn.edu
}
\altaffiltext{2}{
Max-Planck-Institut f\"ur Astronomie, K\"onigstuhl 17,
D-69117 Heidelberg, Germany;
cannon@mpia.de
}

\begin{abstract}			
We report 4.5~\micron\ luminosities for 27 nearby ($D \la$ 5~Mpc) 
dwarf irregular galaxies measured with the {\em Spitzer\/} Infrared
Array Camera.
%
We have constructed the 4.5~\micron\ luminosity-metallicity (\lz)
relation for 25 dwarf galaxies with secure distance and interstellar
medium oxygen abundance measurements.
The 4.5~\micron\ \lz\ relation is
12$+$log(O/H) = (5.78 $\pm$ 0.21) $+$ ($-$0.122 $\pm$ 0.012) $M_{[4.5]}$,
where $M_{[4.5]}$ is the absolute magnitude at 4.5~\micron.
The dispersion in the near-infrared \lz\ relation is
smaller than the corresponding dispersion in the optical \lz\ relation.
The subsequently derived stellar mass-metallicity (\mz) relation is 
12$+$log(O/H) = (5.65 $\pm$ 0.23) $+$ (0.298 $\pm$ 0.030) log~\mstar,
and extends the SDSS \mz\ relation to lower mass by about 2.5~dex.
We find that the dispersion in the \mz\ relation is similar over five
orders of magnitude in stellar mass, and that the relationship between
stellar mass and interstellar medium metallicity is similarly tight
from high-mass to low-mass systems.
We find a larger scatter at low mass in the relation between 
effective yield and total baryonic mass.
In fact, there are a few dwarf galaxies with large yields,
which is difficult to explain if galactic winds are ubiquitous
in dwarf galaxies.
The low scatter in the \lz\ and \mz\ relationships are difficult to
understand if galactic superwinds or blowout are responsible for the
low metallicities at low mass or luminosity.  
Naively, one would expect an ever increasing scatter at lower masses,
which is not observed. 
\end{abstract}


\section{Introduction}                  
\label{sec_intro}

The luminosity-metallicity (\lz) relation for gas-rich star-forming
galaxies is a robust relationship which holds over ten magnitudes
in galaxy optical luminosity 
(e.g., \citealp{zkh94,tremonti04}).
The physical basis for this relation has generally been attributed
to a mass-metallicity relation for galaxies.
The \lz\ relation has also become an important tool to examine 
the time-evolution of past chemical enrichment and the (stellar) 
mass-metallicity (\mz) relation for more massive galaxies at distant
epochs, where emission lines at rest-frame optical wavelengths are
redshifted into the near-infrared; see, e.g.,
\cite{kz99,chip03,liang04,kk04,shapley04,gallazzi05,hoyos05,maier05,savaglio05,shapley05,erb06,lamareille06,liang06}.
The growing body of observations at intermediate- and high-redshifts 
has led to work on a variety of galaxy formation models and
their subsequent predictions for \lz\ and \mz\ relations
(e.g., \citealp{sp99,bb00,tamura01,dw03,delucia04,savaglio05,tissera05}).

For nearby dwarf irregular and other star-forming dwarf galaxies, the
corresponding \lz\ relation has traditionally been studied at optical
wavelengths (e.g., 
\citealp{lequeux79,skh89,rm95,sbk97,pilyugin01,garnett02,hgo03,lee03south,lee03field,lee03virgo,pvc04,vzh06,vanzee06}).
If systematic uncertainties are minimized, the resulting residuals in 
the \lz\ relation can be used to examine effects of galaxy evolution
(e.g., \citealp{chip03}).
However, dispersion in the optical \lz\ relation is thought to arise 
from variations in the stellar mass-to-light ratios, which are
caused by variations in the current star formation rate (SFR) among
galaxies (see, e.g., \citealp{belldejong01,tumlinson06}).
To minimize the effects of these variations, the \lz\ relation can
be determined at near-infrared (NIR; 1--5 \micron) wavelengths, where
the dominant emission arises from the stellar populations.

Recent ground-based efforts to determine \lz\ and \mz\ relations for
low-mass systems in the NIR have been constructed and described by, e.g., 
\cite{pg03}, \cite{jlee04}, \cite{salzer05}, \cite{saviane05},
\cite{oliveira06}, and \cite{rosenberg06}.
The sensitivity of the {\em Spitzer Space Telescope} provides an
excellent opportunity to observe total emission from the 
stellar populations in nearby dwarf galaxies, because most of these
systems are of intrinsically low surface brightness, and are therefore
highly challenging to observe in the NIR from the ground
(e.g., \citealp{vaduvescu05}).
The 4.5~\micron\ luminosities are dominated by emission from
stellar photospheres (e.g., \citealp{lu03,dale05,galliano05}), and
are less sensitive to bursts of star formation.
Observations at 4.5~\micron\ were chosen also because the passband was
designed to avoid emission from polycyclic aromatic hydrocarbons
(PAHs; e.g., \citealp{lp84,allamandola85}).
PAH emission does not appear to be significant within lower-luminosity
metal-poor galaxies (e.g., \citealp{houck04,engelbracht05,hogg05}).
Although emission from hot dust might be significant in disk
galaxies at 4.5~\micron\ (e.g., \citealp{roussel05}), our 8~\micron\
observations have shown that the contributions from PAHs and warm dust
are not significant in our sample of low-mass dwarf galaxies
\citep{jackson06}.

 
In light of the \mz\ relation for massive galaxies determined by
\cite{tremonti04}, we consider nearby dwarf galaxies, which extend the
mass range down by roughly 2.5~dex. 
This allows us to examine the \mz\ relation over five orders of
magnitude in stellar mass.
To construct the \lz\ and \mz\ relations at 4.5~\micron, we have
chosen nearby star-forming dwarf irregular galaxies that have
distances measured from stellar constituents, and oxygen abundances
derived from high-quality spectra (namely, \othreea\ measurements).
We have assembled a sample of 27 dwarf galaxies at
distances below about 5~Mpc; some properties of these dwarf galaxies
are listed in Table~\ref{table_sample}.

The remainder of this paper proceeds as follows.
Those who wish to skip the details should go directly to
\S~\ref{sec_discuss}, where we interpret our key results in the
context of the findings by \cite{tremonti04}.
Observations and reductions of the data are presented in
\S~\ref{sec_obs}.
We describe the measurements of flux densities and absolute magnitudes
of dwarf galaxies at 4.5~\micron, and the corresponding NIR \lz\ and
\mz\ relations in \S\S~\ref{sec_lz} and \ref{sec_mz}. 
We have also gathered from the literature other \lz\ and \mz\
relations for local and distant star-forming galaxies.
As all of the galaxies in our sample have direct measurements of their
\hi\ gas, we also examine relations with gas-to-stellar mass ratio and
\hi\ rotation velocity in \S\S~\ref{sec_gasstar} and \ref{sec_vrot},
respectively.
We give additional discussion and a summary in \S\S~\ref{sec_discuss}
and \ref{sec_concl}, respectively.
For the present discussion, we use the notation
[O/H] = log(O/H) $-$ log(O/H)$_{\odot}$, where the solar value
of the oxygen abundance is 12$+$log(O/H) = 8.66 
\citep{asplund04,melendez04}.
We adopt $Z_{\odot} = 0.0126$ as the revised solar mass fraction in
the form of metals \citep{asplund04}.

\section{Observations and Reductions}
\label{sec_obs}	
	
Data were acquired with the Infrared Array Camera (IRAC;
\citealp{fazio04}) on the {\em Spitzer Space Telescope\/} 
\citep{werner04} during the first year of nominal operations 
from 2003 December to 2004 December. 
In Guaranteed Time Observation (GTO) program 128 (P.I. R.D. Gehrz), ten
galaxies were imaged in channels 2 (4.5~\micron) and 4 (8~\micron).
The Astronomical Observation Request Key (AORKEY) for each galaxy is
listed in Table~\ref{table_sample}. 
We performed five dither positions per pointing with exposure times of
200 seconds each, except for NGC~55 and WLM which had exposure times of
100 seconds per dither position. 
A single $5\arcmin\,\times\,5\arcmin$ pointing was used for all
targets, except for IC~1613, NGC~55, NGC~3109, and WLM,
which were mosaicked in array coordinates using
a $2 \times 2$ ($10\arcmin\,\times\,10\arcmin$), 
a $1 \times 2$ ($5\arcmin\,\times\,10\arcmin$), and 
a $1 \times 3$ ($5\arcmin\,\times\,15\arcmin$)
pointing grid, respectively. 
NGC~55 was mosaicked in sky coordinates, 
resulting in a $15\arcmin\,\times\,10\arcmin$ field-of-view.
Using the MOPEX\footnote{
MOPEX is available from the {\em Spitzer\/} Science Center at
http://ssc.spitzer.caltech.edu/postbcd/.} 
reduction package, the final mosaics were created 
as described in \cite{jackson06}. 
These images contained the appropriate astrometry in Epoch J2000
coordinates and were flux-calibrated in units of MJy~sr$^{-1}$.
Postage-stamp images at 4.5~\micron\ of ten galaxies from 
program GTO~128 are shown in Fig.~\ref{fig_stamps};
note that the galaxies are displayed with the same isophotal 
scaling to emphasize the large range in surface brightness.

Total photometry was obtained with the application of
elliptical apertures, and adjusting the dynamic range to
show the spatial extent of the emission at 4.5~\micron.
We also retrieved optical images from NED and the Local Group 
Survey\footnote{
More about the Local Group Survey is found at 
\url{http://www.lowell.edu/users/massey/lgsurvey.html}.
} 
\citep{massey_lgs},
and we ``stretched'' these images visually (in intensity scale) in
order to discern the outer limit of the optical extent and to compare
the spatial extents of the optical and NIR luminosities.
Elliptical apertures were constructed to optimize the amount of light 
measured for each target galaxy.
Six to seven circular apertures in the immediate vicinity of the  
elliptical aperture were used to sample and determine the mean 
background level surrounding the galaxy. 
This method benefits from the sampling of both the sky and the
foreground stellar contamination.
An example is shown in Fig.~\ref{fig_gxyphot}.
We used the IRAF\footnote{
IRAF is distributed by the National Optical Astronomical
Observatories, which are operated by the Associated Universities for
Research in Astronomy, Inc., under cooperative agreement with the
National Science Foundation.
}
task {\em polyphot\/} to perform the aperture photometry and measure
flux densities. 
A variety of tests were performed to mask and remove negative pixels,
bright foreground stars, and bright pixels caused by ``bleeding''
from bright stars.
Additional experiments (varying the size of the elliptical aperture;
increasing the number of circular ``background'' apertures) were used
to test our final photometry results.
The subsequent range in derived luminosities falls well
within the quoted uncertainties for the distance modulus and
the total errors for IRAC flux measurements (see below).

We have also added from the {\em Spitzer\/} public data archive
eleven dwarf galaxies from the SINGS Legacy program \citep{sings},   
one dwarf galaxy from program GTO~59 (P.I. G.~Rieke), and
five dwarf galaxies from program GTO~69 (P.I. G.~Fazio).
To ensure homogeneity in the derived absolute magnitudes
at 4.5~\micron, we have performed the same procedure
described above to determine total fluxes and luminosities.
As shown in Fig.~\ref{fig_dale},
our measured flux densities for the selected SINGS galaxies
are consistent with those reported by \cite{dale05}.

We adopt 10\% as the error in the total flux, assuming this is the
maximum error of the absolute calibration in channel~2 \citep{hora04}.
Various image-based corrections (array location dependent
photometric correction, pixel-phase correction)
in channel~2 are well below the uncertainty in the absolute
calibration \citep{reach05}, and are not included in this paper.
We have applied a correction to account for extended
surface brightness by multiplying measured channel~2 fluxes by a
factor of 0.937 \citep{reach05} to obtain the appropriate scaled
results. 
With our measured flux densities at 4.5~\micron, we derived absolute
magnitudes, $M_{[4.5]}$, by using the adopted zero-magnitude flux
density, $F^0_{[4.5]} = 179.7 \pm 2.6$ Jy \citep{reach05}.
We have also assumed zero extinction at 4.5~\micron\
(e.g., \citealp{lutz99,indeb05}), and negligible contribution by 
diffuse \hi\ Brackett-$\alpha$ emission to the overall galaxy flux
(e.g., \citealp{lu03,lu04}).

A comparison between $B$ and 4.5~\micron\ luminosities is
shown in Fig.~\ref{fig_lcomp}, where $B$ luminosities corrected
for Galactic extinction are taken from \cite{kara04}.
For the remainder of this paper, we will refer to IRAC channel~2 
luminosities as ``[4.5]''.
Galaxy luminosities at 4.5 \micron\ are also plotted against
\btwo\ color; no correlation is evident.
The mean \btwo\ color for our sample in Table~\ref{table_sample} is
$2.08 \pm 0.59$, which is consistent with the range of observed 
$B\!-\!K$ colors for a sample of nearby late-type dwarf galaxies in 
2MASS \citep{jarrett03}.

\section{The Near-Infrared Luminosity-Metallicity Relation}
\label{sec_lz}

Twenty-five dwarf galaxies in our sample have reported oxygen
abundances, and 21 dwarf galaxies have oxygen abundances derived
from \othreea\ measurements (Table~\ref{table_sample}).
In the absence of \othreea, oxygen abundances for DDO~53, Ho~I,
and NGC~3109 were derived using the \cite{mcgaugh91} bright-line
calibration\footnote{
Analytical expressions for the calibration are found in \cite{chip99}.
Measurements of [\ion{N}{2}]/[\ion{O}{2}] ($\la 0.1$) break the
double-valued degeneracy and point to the ``lower branch,'' which is 
appropriate for dwarf galaxies; 
see \cite{ls04} for additional discussion.
}. 
We adopt for Peg~DIG the oxygen abundance derived by \cite{sbk97}.
We have assumed oxygen abundances are spatially homogeneous in dwarf
irregular galaxies (e.g., \citealp{ks97,chip97,ls04,lsv05,lsv06}).

To obtain fits between absolute magnitude and oxygen abundances
for our sample of dwarf galaxies in $B$ and at 4.5~\micron,
we used the ordinary least squares bisector method 
(see \citealp{linreg3}, and references therein). 
Specifically, these are bivariate least-squares fits with
the bisector method incorporating heteroscedastic errors
(in both luminosity and oxygen abundance).
The two fits are expressed as
\begin{eqnarray}
12+{\rm log(O/H)} & = &
(5.94 \pm 0.27) + (-0.128 \pm 0.017) \, M_B \\
& = &
(5.78 \pm 0.21) + (-0.122 \pm 0.012) \, M_{[4.5]},
\label{eqn_zlumfits}
\end{eqnarray}
where
$M_B$ is the absolute magnitude in $B$, corrected
for Galactic extinction \citep{kara04},
and
$M_{[4.5]}$ is the absolute magnitude at 4.5 \micron, respectively.
The fits in $B$ and [4.5] are shown in Fig.~\ref{fig_oxyll}.
The slopes, intercepts, and their corresponding errors are computed
with the bi-variate correlated errors and
scatter regression code\footnote{
The code is available from STATCODES: 
\url{http://www.astrostatistics.psu.edu/statcodes/}.
},
where we have assumed uncorrelated errors and set the covariance
to zero.
The resulting errors incorporate uncertainties associated with the
data points being fitted.
All of the fits below use the same technique, and we refer to this
procedure as the ``bisector'' method\footnote{
For each fit, we perform 10000 ``bootstrap'' simulations to converge
onto a solution with the appropriate errors.
}. 
The fit at $B$ is consistent with similar fits obtained by, e.g.,
\cite{lee03field}, \cite{vzh06}, and \cite{vanzee06}.
The NIR slope is marginally smaller than the optical slope, although
the derived slopes and intercepts for both \lz\ relations are
consistent with each other. 

Variations in the stellar mass-to-light ratio can be responsible for
some or all of the scatter in the \lz\ relation.
Moreover, the optical luminosity of a fainter galaxy is affected
more by a burst of new star formation than the luminosity of a
brighter galaxy. 
\cite{sbk97} proposed a comparison of the residuals in luminosity 
against color to test whether variations in the stellar mass-to-light
ratio were the main cause of the scatter.
A limited set of data at the time were plotted against optical $B-V$
color, and were suggestive of a correlation between the residuals
in $B$ luminosity versus $B-V$.

We show now that the scatter in the \lz\ relation diminishes
when luminosities are measured at longer wavelengths.
Residuals in oxygen abundance are plotted against $B$ and
[4.5] absolute magnitude in the top and bottom panels, respectively,
of Fig.~\ref{fig_offmag}.
The dispersions in the optical and NIR \lz\ relations
are 0.161~dex and 0.122~dex, respectively.
The scatter in the NIR \lz\ relation is reduced compared 
to the optical \lz\ relation.
The dispersion in the \lz\ relation was {\em expected\/} 
to decrease at longer wavelengths, as NIR luminosities are less
sensitive to extinction by dust and variations in the present-day
SFR.

The smaller slope and the reduced dispersion in the NIR are in
agreement with the results for star-forming galaxies in the field
reported by \cite{salzer05}.
Higher-luminosity and higher-mass galaxies contain more metals
and more dust (e.g., \citealp{rosenberg06}); these galaxies would be
optically underluminous, which would yield a steeper slope in the
optical \lz.
The effects of absorption by dust vary from galaxy to galaxy, which
introduces greater overall scatter in the optical \lz.
Absorption by dust is mitigated in the NIR, which produces
an \lz\ relation with a shallower slope and less scatter.
For our present sample of dwarf galaxies, we have assumed that dust
is less important (see \S~\ref{sec_intro}) than the variations in
stellar mass-to-light ratios.

We now examine residuals in oxygen abundance and luminosity 
against $B-[4.5]$ color. 
The results are shown in Fig.~\ref{fig_offcol}.
Residuals in oxygen abundance from the $B$ \lz\ relation are
correlated with color in the sense that redder galaxies have
larger abundances compared to the best fit. 
On the other hand, there is very little to no trend with
color for the residuals from the [4.5] \lz\ relation.
Residuals in magnitude from the optical and infrared \lz\ relations
exhibit similar behavior as above; redder galaxies are fainter in $B$
and no trend is found at 4.5~\micron.
The plots in Fig.~\ref{fig_offcol} clearly show that the dispersion is
clearly reduced at NIR wavelengths. 

We compare in Table~\ref{table_lzfits} the 4.5~\micron\ \lz\ relation
with recent optical and NIR \lz\ relations from the literature.
The slopes from our optical and NIR relations are smaller than values
derived from similar relations for other samples of nearby dwarf
galaxies.
In addition to listing the slope and intercept values for each
\lz\ relation, we have also computed the oxygen abundance from
the specified relation at a fiducial galaxy luminosity equal
to $M_{\lambda} = -16$.
The key result from the comparison is that our NIR 
\lz\ relation exhibits the smallest dispersion.
%

\section{Stellar Mass-Metallicity Relation}
\label{sec_mz}

The NIR \lz\ relation reflects a fundamental relation between
underlying stellar mass and metallicity, and thus,
NIR luminosities of dwarf galaxies can directly be related to the
mass of their stellar populations
(e.g., \citealp{vaduvescu05}).
First, to derive the 4.5~\micron\ luminosity in solar units, we have 
assumed that the absolute magnitude of the Sun is $+3.33$
in $K$ \citep{belldejong01}, and that the ground-based $K\!-\!M$ 
color\footnote{
We assume that the effective wavelength of $M$ is 
4.8~\micron\ \citep{crl85}.
}
is about 0 for a G2 dwarf star (e.g., \citealp{bb88}).
Thus, we assume that the absolute magnitude of the Sun at 4.5~\micron\
is $\simeq +3.3$.
Even if the NIR luminosity is dominated by giant stars,
the most extreme \ktwo\ colors are between $+0.1$ and $-0.1$. 
%
%
As we show below, a non-zero \ktwo\ color introduces a zero-point
offset in the \mz\ relation.

We consider an appropriate stellar mass-to-NIR-light ratio.
We first assumed a constant stellar mass-to-[4.5]-light ratio 
of 0.5, as motivated by models for the formation of disk 
galaxies (e.g., \citealp{vandenbosch02}).
This value of the stellar mass-to-light ratio is assumed
to be independent of color.
The resulting fit to the stellar mass-metallicity relation is 
12$+$log(O/H) = $(5.26 \pm 0.27) + (0.332 \pm 0.033)$ log \mstar,
and the resulting dispersion is 0.125~dex.
However, stellar mass-to-light ratios can vary by a factor of 
$\sim 2$, even at NIR wavelengths (e.g., \citealp{belldejong01}).
Fortunately, stellar mass-to-NIR light ratios are not strongly
dependent upon the age of the dominant stellar population.

For the next step, we derived stellar mass-to-light ratios by
accounting for color variations.
We used the models by \citet[][their Tables 3 \& 4]{belldejong01},
where the authors developed methods to derive better estimates of
stellar mass in disk galaxies and to determine the baryonic
Tully-Fisher relation.
For each model in \cite{belldejong01}, we derived a mass-to-light
ratio as a linear function of \bk\ color, and computed the stellar
mass for each galaxy.
Using the color-based functions for \bv\ and \vk,
we applied a simple linear combination to obtain the stellar
mass-to-light ratio as a linear function of \bk\ color.
We write 
\begin{equation}
\log\,(\mstar/L_K) = a^{\prime} + b^{\prime}\,(\bk),
\label{eqn_mlk}
\end{equation}
where $a^{\prime}$ and $b^{\prime}$ are given in
Table~\ref{table_masses}, and 
$\bk = (\btwo)-(\ktwo)$ using our measured \btwo\ colors.
For a small sample of late-type galaxies, \cite{pahre04} determined 
\ktwo\ in the range between $-0.1$ and $+0.4$.
We adopt here $\ktwo = +0.2$, and subsequent stellar masses 
are 0.15 to 0.16~dex smaller than stellar masses derived
with $\ktwo = 0$.

We obtain stellar masses with
\begin{eqnarray}
\log\,\mstar & = & \log\,(\mstar/L_K) + \log\,L_K \nonumber \\
& = & \log\,(\mstar/L_K) + \left[\log\,L_{[4.5]} - 0.4\,(\ktwo)\right],
\label{eqn_getmstar}
\end{eqnarray}
with the stellar mass-to-light ratio defined above.
With this empirical prescription for stellar masses, we 
subsequently performed a \mz\ fit to each model.
The results are shown in Table~\ref{table_masses}.
Although stellar masses for a given galaxy can vary by as much
as $\sim 0.5$~dex from model-to-model, the subsequent \mz\ fits 
spanning a range of about four decades in stellar mass are similar
for all models.
The \mz\ slopes for models at subsolar metallicity are marginally
smaller than the scaled Salpeter stellar initial mass function (IMF)
models at solar metallicity.

For this work, we used model~8 in Table~\ref{table_masses} 
(Salpeter IMF, subsolar metallicity)
and Equations~(\ref{eqn_mlk}) and (\ref{eqn_getmstar}) 
to derive initial stellar masses.
We then subtracted 0.25~dex from all of the stellar masses
to scale the masses from the \cite{salpeter55} IMF to the
\cite{chabrier03} IMF.
\cite{tremonti04} used the \cite{kroupa01} IMF to derive
stellar masses, which are about 0.30~dex smaller than masses
derived with the Salpeter IMF;
\cite{bell03} provide additional discussion on using
different IMFs to derive stellar masses in late-type galaxies.
We adopt the errors in [4.5] absolute magnitude as the errors
in the logarithm of the stellar mass.
We note that our derived stellar mass for NGC~6822,
log (\mstar/\msun) = 8.10, is consistent with
the value (8.24) derived from an independent model describing the
photometric and chemical evolution \citep{carigi06}.

The final \mz\ relation is expressed as
\begin{equation}
12+{\rm log(O/H)} =
(5.65 \pm 0.23) + (0.298 \pm 0.030) \, \log \mstar.
\label{eqn_zmstar}
\end{equation}
The dispersion is 0.117~dex.
In Fig.~\ref{fig_zmstar}, we show the data and the fit expressed by
Equation~(\ref{eqn_zmstar}) in panel (a). 
We find an excellent correlation between oxygen abundance and stellar
mass; the Pearson correlation coefficient is $r = +0.90$.
%
%
No trend is seen in a plot of the residuals in oxygen abundance
from the \mz\ relation against stellar mass in panel (b).
%

We compare our \mz\ relation with the relation recently
derived for massive galaxies.
\cite{tremonti04} derived the \mz\ relation for over 53000
massive star-forming galaxies at redshifts $z \la 0.1$ from the Sloan
Digital Sky Survey (SDSS).
In Fig.~\ref{fig_sdss}, we show the full range in both axes, spanning
over two orders of magnitude in oxygen abundance, and five orders of
magnitude in stellar mass.
The data presented here extend the SDSS \mz\ relation to lower
masses by roughly 2.5~dex.
The dispersion for our \mz\ relation is comparable 
to that of the SDSS \mz\ relation ($\approx$ 0.1~dex). 
If dwarf galaxies have insufficient total masses to retain ejecta
from supernovae, we expect that the dispersions in both \lz\ and \mz\
relations should increase with decreasing luminosity and mass.
We will discuss the implications of our \lz\ and \mz\
results in \S~\ref{sec_discuss}.
%

The polynomial fit to the SDSS galaxies is effective in
the mass range between $10^{8.5}$ and $10^{11.5}$ \msun, whereas
our relation overlaps the SDSS relation in the mass range between
$10^{8.5}$ and $10^{9.5}$ \msun. 
At these stellar masses, SDSS galaxies are about 0.2--0.3~dex more
metal-rich than the present sample of dwarf galaxies.
We describe here the most likely reasons for the offset; 
see also \cite{savaglio05}.
First, SDSS fiber spectra preferentially sample the inner 
$\approx$ 25\% of galaxies \citep{tremonti04}.
Given that nearby spirals exhibit radial metallicity gradients
(e.g., \citealp{zkh94}), derived abundances from SDSS spectra might be
too high for a number of the galaxies.
Second, oxygen abundances for the SDSS data were derived using
bright-line methods in the absence of the temperature-sensitive
\othreea\ emission line.
Recent work with nearby spiral galaxies has shown that bright-line
oxygen abundances overestimate the true oxygen abundances by roughly
0.2 to 0.3 dex at metallicities solar and above 
(e.g., \citealp{stasinska02,kbg03,bgk04}).
These effects together can account for the offset between the
present sample of dwarf galaxies and the SDSS sample of higher-mass
galaxies.

Fig.~\ref{fig_sdss} suggests two intriguing possibilities:
(a) over a large range in mass ($\simeq 5.5$ dex) there is a single
and continuous \mz\ relation which turns over or flattens at high galaxy
mass,
or
(b) the \mz\ relations describing dwarf and giant galaxies may
have different slopes; this possibility has also been suggested by
e.g., \cite{salzer05,rosenberg06}.
It will be very interesting to investigate whether slopes of the \lz\
and \mz\ relations exhibit a ``break'' at some characteristic luminosity
and mass, respectively\footnote{
Examination of the latest SDSS data 
($\sim$ 80000 objects in DR4; C. Tremonti, personal communication) 
shows that the subsequent \mz\ fit is effective to another 0.5 dex
lower in mass to log (\mstar/\msun) $\sim 8$.
Despite the different methods by which stellar masses and oxygen
abundances have been determined, there appears to be good agreement at
log (\mstar/\msun) $\sim 8$ between the upper-end of our \mz\ relation
and the lower-end of the SDSS \mz\ relation.
}.


In Table~\ref{table_mzfits}, we list the \mz\ relation
from the present work, as well as other relations from the literature.
With increasing numbers of intermediate- and high-redshift galaxies
with measured luminosities and abundances, the interpretation of the
\mz\ relation for massive systems at distant epochs will
depend upon the specific bright-line calibration used to derive oxygen
abundances.
Although selection criteria and the methods used to derive
masses and metallicities may vary, comparisons which begin to
tie relations between high-mass and low-mass galaxies and
between local and distant galaxies may help illustrate 
processes acting on galaxies over a range in mass and redshift.

\section{Relations with Gas-to-Stellar Mass Ratio}
\label{sec_gasstar}

We consider the gas-to-stellar mass ratio, which is representative of
the baryonic gas fraction 
(e.g., \citealp{garnett02,lee03field,scm03,vzh06}).
These parameters reveal important information about the conversion of
gas into stars and metals, and about possible gas flows.
Assuming zero gas flows and instantaneous recycling, the closed-box
model of chemical evolution predicts the metallicity by mass $Z$
\citep{schmidt63,ss72}
\begin{equation}
Z = y \, \ln \, \mu^{-1},
\label{eqn_chemev1}
\end{equation}
where $y$ is the yield and 
$\mu$ = $M_{\rm gas}/(M_{\rm gas} + \mstar)$ is the 
fraction of total baryonic mass in the form of gas.
From the equation above, we can derive the oxygen abundance
by number 
\begin{equation}
12 + \log({\rm O/H}) =
12 + \log(y_{\rm O}/\zeta) + 
\log \left( \ln \left[ 1 + (M_{\rm gas}/M_{\ast})^{-1} 
\right] \right),
\label{eqn_chemev2}
\end{equation}
where 
$y_{\rm O}$ is the oxygen yield by mass, 
$\zeta = 11.728$ is the factor to convert abundance by mass to
abundance by number, 
$M_{\rm gas}$ = 1.36 $M_{\rm HI}$ is the total gas mass
(which includes helium),
and 
$M_{\rm gas}/M_{\ast}$ is the gas-to-stellar mass ratio,
which is independent of distance.

All of the galaxies in the present sample have direct measures of
their \hi\ gas content.
We have adopted 21-cm flux integrals from the catalog by \cite{kara04}.
However, we have not scaled the \hi\ mass to the spatial extent of the
optical or NIR luminosity (re. stellar mass)\footnote{
The proper interpretation of effective yields
with respect to the ``inner'' and ``outer'' gas relative
to the optical extent depends very strongly on the assumed metallicity
of the outer gas, which is not known or well constrained in many
nearby dwarf galaxies; however, see, e.g., 
\cite{skh89,ks01,garnett02} for additional discussion.
A detailed comparison between ``non-scaled'' and ``scaled'' \hi\ 
content relative to the optical extent is outside the present focus 
of this paper.
}.
The 21-cm flux integrals and the derived gas-to-stellar mass fractions
are listed in Table~\ref{table_gas}.

The results are shown in Fig.~\ref{fig_gasstar}.
In panel (a), we have plotted oxygen abundances
against the gas-to-stellar mass ratio, and
the subsequent least-squares fit with the bisector method.
The fit is shown as a solid line, and is expressed as
\begin{equation}
12 + \log({\rm O/H}) = 
(8.21 \pm 0.07) + (-0.435 \pm 0.085) \log(M_{\rm gas}/\mstar).
\label{eqn_gasstar}
\end{equation}
The dispersion is 0.24~dex.
In the center panel (b), we have plotted outflow models
as variations of the closed-box model; we assume negligible
inflow of metal-poor gas.
The purple and blue set of curves, respectively, describe 
$y_O = 0.01$ \citep{tremonti04} and
$y_O = 2 \times 10^{-3}$ \citep{lee03field,vzh06}. 
For each set of curves, the outflow rates from right to left
are 0, 5, 10, 25, and 50 times the star-formation rate.
These results suggest reduced yields and/or significant
outflow rates, which previous authors (e.g.,
\citealp{garnett02,vzh06}) have also indicated.

In Fig.~\ref{fig_gasstar}c, we have plotted the \btwo\ color
against the gas-to-stellar mass ratio, where there is a very
clear correlation.
We first point out three outliers in the data.
At \btwo\ $\approx +1$ and log ($M_{\rm gas}/$\mstar) $\approx -0.4$,
NGC~1569 is unusually blue for its gas-to-stellar mass ratio, as
this dwarf galaxy is known to have just undergone a very strong burst 
of star formation (e.g., \citealp{israel88,waller91}).
At the other end of the plot (\btwo\ $\approx +2$,  
log ($M_{\rm gas}/$\mstar) $\approx +1.5$) are DDO~154 and Ho~IX.
DDO~154 is known to have a very large \hi\ halo \citep{cb89,cp98}.
Because Ho~IX is situated (in projection) very close to the spiral
galaxy M~81, the \hi\ for Ho~IX is most likely a part of the overall
gas distribution for M~81.

The bisector fit to the data (excluding DDO~165, Ho~IX, and
NGC~1569) is shown as a solid line, and is expressed as
\begin{equation}
(B-[4.5]) = 
(2.79 \pm 0.11) + (-0.89 \pm 0.13)\, \log(M_{\rm gas}/\mstar),
\label{eqn_gascolor}
\end{equation}
and the dispersion in color is 0.32~mag.  
The fit also implies that 
$L_B/L_{[4.5]} \propto (M_{\rm gas}/\mstar)^{0.36}$.
The strong correlation may simply reflect a more dominant process of
the conversion of gas into stars.
Galaxies with lower gas-to-stellar mass ratios have already depleted
their fuel, and appear redder when present-day star formation no
longer becomes active.

We consider possible departures from the closed-box model by 
inverting Equation~(\ref{eqn_chemev1}) to obtain the effective yield,
$y_{\rm eff} = Z/\ln(\mu^{-1})$.
The effective yield is determined simply from the measured
oxygen abundances and baryonic gas fractions, and is 
equal to the true yield when the closed-box assumption applies. 
Effective yields which are smaller than the true yield suggest
either outflow or infall of metal-poor gas
(e.g., \citealp{edmunds90,dalcanton06,vzh06}).
We have listed derived values of the effective yield in
Table~\ref{table_gas}.

In Fig.~\ref{fig_yield}a, we plot the effective yield as a 
function of the total baryonic mass, following Fig.~8 in
\cite{tremonti04}; \cite{garnett02} also obtained a similar plot for a
sample of nearby spiral and irregular galaxies. 
The empirical fit obtained by \citep{tremonti04} is based upon the
premise that galactic winds were responsible for metals loss;
see also \citealp{veilleux05} for additional discussion.
From the present sample, there are dwarf galaxies with effective
yields about a factor of ten lower than the ``asymptotic'' yield 
(log $y_{\rm eff} \approx -1.98$) for more massive galaxies in the SDSS.
We have also marked for comparison the solar oxygen mass fraction 
(log $Z_{{\rm O},\odot} = -2.27$, \citealp{asplund04,melendez04}) and 
the effective yield for the nearby spiral M51
(log $y_{\rm eff} = -2.49$, \citealp{bgk04})\footnote{
For a sample of \hii\ regions in M51, \cite{bgk04} measured 
electron temperatures from \ntwot\ measurements, and derived chemical
abundances.
The subsequent effective yield for M51 was found to be lower by a
factor of four than previous estimates.
}.
Accounting for the $\sim$~0.2--0.3 dex offset described above
in \S~\ref{sec_mz} lowers the ``asymptotic'' yield in better
agreement with the solar mass fraction and the effective
yield for M51.
We note that these values are also comparable to the range
of effective yields for dwarf galaxies 
(e.g., log $y_{\rm eff} \simeq$ $-2.7$ to $-2.1$) previously
reported by, e.g., \citealp{lee03field,vzh06}).

Although there may be a trend between yield and baryonic mass for the
low-mass galaxies, there is however large scatter, which is about
1.5~dex at $\sim$ $10^{8.5}$~\msun.
Three galaxies with the largest effective yields
(log~$y_{\rm eff} \ga -2$) in our sample are DDO~154, NGC~3109, and
Sextans~A.
However, their measured luminosities and abundances are in good
agreement with other galaxies at comparable abundances and
luminosities, respectively.
%

\cite{vzh06} observed a trend between effective yield and
H~I-gas-to-$B$-light ratio for a sample of nearby UGC dwarf galaxies,
where the H~I-gas-to-$B$-light ratio was a measure of the 
gas-to-stellar mass ratio.
These authors ruled out infall of metal-poor gas as a mechanism to
decrease effective yields (see also \citealp{dalcanton06}).
In Fig.~\ref{fig_yield}b, we plot the effective yield as a function
of the gas-to-stellar mass ratio.
Gas-rich systems eventually reach the ``asymptotic'' yield,
while gas-poor systems have lower effective yields.
The effective yield is correlated with the gas-to-stellar mass ratio,
and the fit is expressed as
\begin{equation}
\log y_{\rm eff} = (-2.75\,\pm\,0.06) + (0.700\,\pm\,0.057)
\log (M_{\rm gas}/\mstar)
\label{eqn_yield}
\end{equation}
with dispersion in log yield equal to 0.22~dex.
In the limit of high gas fraction, the closed-box model
predicts that the slope in equation above is equal to one.

\section{Relations with Rotation Velocity}
\label{sec_vrot}

For a sample of dwarf irregular and spiral galaxies, \cite{garnett02}
found a relationship between metallicity and rotation velocity with an
apparent ``break'' in the relation at 120 km~s$^{-1}$.
\cite{pvc04} peformed a homogeneous analysis using his
bright-line calibration, and found a similar break at 80 km~s$^{-1}$.
We plot various quantities against \hi\ rotational velocity in
Fig.~\ref{fig_vrot}, where the range of velocities for the present 
sample is between 10 and 80 km~s$^{-1}$; see Table~\ref{table_sample}.
As an alternative to velocity on a linear scale, we plot
the {\em logarithm\/} of the velocity.

In Fig.~\ref{fig_vrot}a, there appears to be a good correlation between
stellar mass and rotation velocity ($r = +0.66$),
although there is a paucity of data at low rotation velocity.
The bisector fit is 
\begin{equation}
\log \mstar = 
(4.14\,\pm\,0.57) \, + \, (2.33\,\pm\,0.36) \log v_{\rm rot},
\label{eqn_vrot}
\end{equation}
and the dispersion in log stellar mass is 0.63~dex.
The fit can be inverted to give 
$v_{\rm rot} \propto \mstar^{0.43}$; 
the numerical exponent here is larger than the predicted value of 0.2
\citep{dw03} by $3.4\sigma$.
The observational result suggests either larger rotation velocities at
a given stellar mass, or smaller stellar masses at a given velocity.

In panels (b) and (c), respectively, we have plotted the
gas-to-stellar mass ratio and the effective yield against
rotation velocity.
Neither panel reveals a strong trend, although the effective yield
appears to rise slightly with rotation velocity.
The rise of the effective yield with rotation velocity has been
discussed by \cite{garnett02}, \cite{pvc04}, and
\cite{dalcanton06}. 
Galaxies with $v_{\rm rot} \ga$ 100 km~s$^{-1}$ have effective yields
approaching that of the asymptotic yield.
Less massive galaxies with $v_{\rm rot} \la$ 100 km~s$^{-1}$ have 
smaller effective yields which are roughly consistent with 
a solar-type yield, as seen above in Fig.~\ref{fig_yield}.
The present work suggests that the effective yield may be roughly
constant, or at the very least, is weakly dependent upon rotation
velocity.
While additional data are required for confirmation, this result
suggests that within the mass range of dwarf galaxies, the effective
yield depends more strongly upon the relative gas fraction than
upon rotation velocity.


\section{Discussion}                    
\label{sec_discuss}

We have shown that the scatter in the optical \lz\ relation is
constant ($\sim$ 0.16~dex) over a range of 11~mag in $B$ luminosity,
and the scatter decreases to 0.12~dex over 8~mag at NIR wavelengths.
Our subsequent stellar mass-metallicity relation extends the dynamic
range of the SDSS \mz\ relation to lower masses by roughly 2.5~dex.
We have also shown that the scatter for the \mz\ relation appears
to be similar over 5.5~orders of magnitude in stellar mass.

Models of galaxy formation have been successful in predicting the
slopes of the \lz\ and \mz\ relations. 
These models generally incorporate feedback, which describes how
effectively massive stars can inject energy into the surrounding
interstellar medium. 
\cite{ds86} predicted the scaling relation, $L \propto Z^{2.7}$,
for dwarf galaxies formed and embedded within dark matter haloes,
although their model is less applicable to dwarf galaxies
which have undergone a low and constant level of star formation
over a Hubble time. 
The derived slopes of the \lz\ relations 
(our Equation~\ref{eqn_zlumfits})
are in approximate agreement with the \cite{ds86} prediction, 
although our [4.5] \lz\ slope is smaller than the predicted value
by about $2\sigma$.
Recently, \cite{dw03} predicted the scaling relation 
$Z \propto \mstar^{0.4}$. 
The slope of the \mz\ relation in Equation~(\ref{eqn_zmstar}) is
smaller than the predicted value by $3.4\,\sigma$.
We note, however, that the \cite{dw03} scaling relation was derived
for stellar iron metallicities under the assumption of constant
[Fe/O], which may not be applicable in dwarf irregular galaxies
(e.g., \citealp{smith02}).

\cite{sp99} compared predicted \lz\ relations among a number of
semi-analytical models. 
Although the details varied with the specific feedback recipe
within the models, the scatter in the \lz\ relations increased with
decreasing total luminosity (their Fig.~18).
Unfortunately, none of the models provided acceptable fits to all of
the known observational parameters, and these authors stated that the
failure of the models is likely caused by a strong difference in
efficiency between ejecting gas and ejecting metals.

Models by \cite{delucia04} produced \mz\ relations consistent with
the SDSS \mz\ relation over a similar range in stellar mass, down
to log (\mstar/\msun) $\sim$ 8.5. 
\cite{robertson05} developed models to describe chemical abundance
patterns observed in the Galactic halo, Local Group dwarf spheroidals,
and Local Group dwarf irregular galaxies.
They incorporated chemical enrichment histories and various gas
accretion histories within cosmologically motivated $\Lambda$CDM
models for the formation of dwarf galaxies and the Galactic stellar
halo. 
Their models were able to reproduce the \cite{dw03} scaling relation
and the observed abundance patterns in nearby dwarf galaxies.

\cite{tremonti04} explained that the relationship between stellar mass
and metallicity was a consequence of the relationship between
effective yield and total baryonic mass.
That is, higher-mass systems are able to convert gas into stars
which produce galaxies with large stellar masses and metallicities;
whereas, lower-mass systems convert less gas into stars, and 
lose their metals as outflows by virtue of the lower total masses.
These authors state: ``\ldots\ It is our view that the strong positive
correlation between effective yield and baryonic mass is most 
naturally explained by the increasing potential barrier that the
metal-laden wind must overcome to achieve `blowout.' \ldots''
In fact, since a number of the low-mass galaxies in our sample 
do not depart significantly from the \lz\ and \mz\ relationships
and have effective yields close to or exceeding the asymptotic value,
it is difficult to explain these relationships as a result of very
disruptive blowout events.

Present-day galactic winds removing material from low-mass galaxies
should have dramatic effects on the properties of such systems.
We consider a prototype dwarf starburst galaxy, NGC~1569,
which is undergoing blowout.
\cite{martin02} showed in their {\em Chandra\/} observations
that most of the oxygen produced in the starburst is in 
the enriched outflow.
However, the results from \cite[][their Table 6]{martin02} and 
\cite{angeretti05} reveal that at most 2\% of the
stellar mass is produced in the burst.
Although the stellar mass ($\approx 10^{8.5}$ \msun) of NGC~1569
is at the upper end of the range for our sample of dwarf galaxies,
an equivalent NGC~1569-like burst would produce a more significant
amount of stars in lower-mass systems.
However, this high level of star formation is not presently seen in
the other low-mass galaxies from our nearby sample, and certainly not
seen in dwarf irregulars within the Local Group (cf. \citealp{mateo98}).


We hypothesize that a less energetic form of metal-enhanced mass loss
than blowouts could explain the small scatter in the \lz\ and \mz\
relationships. 
Perhaps the slow loss of metals from an x-ray corona which is fed by
individual supernova is sufficient to explain the observed
relationships.  
Since this mechanism is an average over many events instead of
depending on a small number of very energetic events, a smaller
scatter would arise naturally.  
A low efficiency of star formation in dwarf irregular galaxies
could account for a gentler form of mass loss.

While a number of the more massive dwarf galaxies in the present
sample have experienced or will experience an outflow event 
(e.g., \citealp{summers03,hartwell04,grimes05,ott05}), 
the challenge remains to explain the uniform scatter in both the
NIR \lz\ relation over a range of about 11~magnitudes in luminosity,
and the \mz\ relation over a range of about 5.5~decades in stellar mass. 
We suggest the following lines of research for the near future.
New generations of models for galaxy formation which
incorporate varying degrees of galaxy outflows should be
able to predict the slope and the scatter in the \lz\ and \mz\
relations.
Large uniform surveys of galaxies are required to examine the
percentage of stars formed in starbursts as a function of galaxy
mass (e.g., \citealp{pg03}).
Results derived from the resolved stars in NGC~1569 (e.g.,
\citealp{angeretti05}) are promising, 
because similar studies will allow us to determine how much stellar
mass has formed in the most recent bursts to a lookback time of a
few Gyr.
Additional NIR imaging and spectroscopy for a large number of dwarf
galaxies within the Local Volume \citep{kara04} can test whether
the scatter in the \lz\ and \mz\ relations is truly comparable to the
scatter in the respective relations for massive galaxies.
With growing interest in understanding further the stages of
galaxy assembly through \lz\ and \mz\ diagnostics, improved models 
are crucial in providing further insights into the evolution of 
local and distant galaxies.

\section{Conclusions}		
\label{sec_concl}		

With {\em Spitzer\/} IRAC imaging, we have derived total
near-infrared luminosities for 27 nearby dwarf irregular galaxies at
4.5~\micron.
The near-infrared \lz\ relation is constructed using 25 of these
galaxies with secure distance and oxygen abundance measurements.
While the slope of the subsequent [4.5] \lz\ relation is slightly
smaller than the corresponding optical relations, the dispersion in
the near-infrared \lz\ relation is reduced compared to the
optical \lz\ relation.
This result agrees with the expectation that variations in stellar
mass-to-light ratios decrease from optical to near-infrared
wavelengths.
We constructed a \mz\ relation, where the stellar masses were derived
using color-based stellar mass-to-light ratios and corrected for
the Chabrier IMF.
The \mz\ relation for dwarf galaxies extends the SDSS \mz\ relation
by another 2.5~decades.
We find that the dispersion in the \mz\ relation is comparable
over 5.5~decades in stellar mass.
The dispersions in the \lz\ and \mz\ relations as well
as the large variation in effective yields are difficult
to understand if galactic superwinds or outflows are responsible
for low metallicities at low mass or luminosity.

\begin{acknowledgements}	

We thank the anonymous referee for their comments.
This work is based in part on observations made with the {\em Spitzer 
Space Telescope}, which is operated by the Jet Propulsion Laboratory, 
California Institute of Technology under NASA contract 1407. 
Support for this work was provided by NASA through contracts
1256406 and 1215746 issued to the University of Minnesota 
by JPL/Caltech.
H.~L. is grateful to 
Eric Bell, Dominik Bomans, Andrew Cole, Chad Engelbracht, Ken Freeman,
Karl Gordon, Valentin Ivanov, Rob Kennicutt, Chip Kobulnicky, Janice Lee, 
Marshall McCall, John Moustakas, Michael Richer, Jessica Rosenberg,
H\'el\'ene Roussel, John Salzer, Ivo Saviane, Rachel Somerville,
Christy Tremonti, and Liese van Zee 
for their comments, suggestions, and words of wisdom.
H.~L. also thanks Julianne Dalcanton, Claudia Mendes de Oliveira, and 
Sonia Temporin for copies of their manuscripts before publication.
H.~L. and E.~D.~S. acknowledge partial support from a NASA LTSARP 
grant NAG~5--9221 and the University of Minnesota.
This research was partially suppported by a grant from the
American Astronomical Society.
Some data were accessed as Guest User, Canadian Astronomy Data Center,
which is operated by the Dominion Astrophysical Observatory for the 
National Research Council of Canada's Herzberg Institute of Astrophysics. 
This research has made use of NASA's Astrophysics Data System, and
of the NASA/IPAC Extragalactic Database (NED), which is operated by
the Jet Propulsion Laboratory, California Institute of Technology,
under contract with the National Aeronautics and Space Administration. 

\end{acknowledgements}


\clearpage	

\begin{table}
\begin{center}
\setlength{\tabcolsep}{1.7mm}
\renewcommand{\arraystretch}{1.3}
\caption{
Properties for 27 nearby ($D \la 5$ Mpc) dwarf irregular galaxies.
\vspace*{2mm}
\label{table_sample}
}
\begin{tabular}{lcccccccccc}
\tableline \tableline
& & & & & & & Nebular \\ 
& & AORKEY & $F_{[4.5]}$ & $m\!-\!M$ & Distance & $M_{[4.5]}$ & 
12$+$log(O/H) & O/H & $B\!-\![4.5]$ & log \mstar \\
Galaxy & Source & (G128) & (mJy) &
(mag) & Refs. & (mag) & (dex) & Refs. & (mag) & (\msun) \\ 
(1) & (2) & (3) & (4) & (5) & (6) & (7) & (8) & (9) & (10) & (11)
\\[1mm]
\tableline
DDO 154 &
        S & \nodata &
	$2.42 \pm 0.24$ & 
        $27.53 \pm 0.10$ & 1 & 
	$-15.28 \pm 0.10$ &
	$7.67 \pm 0.05$ & 1,2 & 
	1.88 & 6.68 
	\\
DDO 165 &
        S & \nodata &
	$9.57 \pm 0.96$ & 
        $28.30 \pm 0.19$ & 2 & 
	$-17.62 \pm 0.19$ &
	\nodata & \nodata & 
	2.46 & 7.73 
	\\
DDO 53 &
        S & \nodata & 
	$2.97 \pm 0.30$ & 
        $27.76 \pm 0.15$ & 2 & 
	$-15.73 \pm 0.15$ &
	$7.62 \pm 0.20$ & 3 & 
	2.36 & 6.98 
	\\
GR 8$\,$\tablenotemark{a} & 
        G128 & 5054464 & 
	$2.95 \pm 0.30$ &
	$26.75 \pm 0.35$ & 3 & 
	$-14.72 \pm 0.35$ & 
	$7.64 \pm 0.04$ & 4 & 
	2.52 & 6.62 
	\\
Ho I &
        S & \nodata & 
	$4.40 \pm 0.44$ & 
	$27.92 \pm 0.26$ & 1 & 
	$-16.32 \pm 0.26$ &
	$7.70 \pm 0.20$ & 5 & 
	1.83 & 7.08 
	\\
Ho II &
        S & \nodata & 
	$42.9 \pm 4.3$ & 
	$27.65 \pm 0.13$ & 1 & 
	$-18.53 \pm 0.13$ &
	$7.93 \pm 0.05$ & 6,7 & 
	1.82 & 7.96 
	\\
Ho IX &
        S & \nodata & 
 	$2.72 \pm 0.27$ &
 	$27.80 \pm 0.08\,$\tablenotemark{b} & \nodata & 
 	$-15.75 \pm 0.10$ &
 	\nodata\tablenotemark{b} & \nodata & 
	2.07 & 6.89 
	\\
IC 1613 &
        G128 & 5051648 & 
	$91.53 \pm 9.2$ &
	$24.22 \pm 0.10$ & 4,5 & 
	$-15.92 \pm 0.10$ & 
	$7.62 \pm 0.05$ & 8 & 
	1.42 & 6.82 
	\\
IC 2574 & 
        S & \nodata & 
	$70.1 \pm 7.0$ &
	$28.02 \pm 0.22$ & 1 & 
	$-19.43 \pm 0.22$ &
	$8.15 \pm 0.11$ & 5,6 & 
	2.09 & 8.39 
	\\
IC 5152 &
        G128 & 5055232 & 
	$149 \pm 15$ &
	$26.58 \pm 0.18$ & 6 & 
	$-18.28 \pm 0.18$ & 
	$7.92 \pm 0.07$ & 8 & 
	2.78 & 8.11 
	\\
Leo A &
        G128 & 5052416 & 
	$8.65 \pm 0.87$ &
	$24.2 \pm 0.2$ & 7 & 
	$-13.34 \pm 0.20$ & 
	$7.35 \pm 0.06$ & 4 & 
	1.84 & 5.89 
	\\
M81 dwB$\,$\tablenotemark{a} &
        S & \nodata & 
	$2.72 \pm 0.27$ &
        $28.62 \pm 0.10$ & 1 & 
	$-16.50 \pm 0.10$ &
	$7.98 \pm 0.22$ & 5 & 
	1.96 & 7.19 
	\\
NGC 1569 &
        G69 & \nodata & 
	$220 \pm 22$ &
	$26.71 \pm 0.60$ & 8 &	
	$-19.36 \pm 0.60$ &
	$8.19 \pm 0.02$ & 9--11 &
	0.93 & 8.07 
	\\
NGC 1705 &
        S & \nodata & 
	$17.4 \pm 1.7$ & 
	$28.54 \pm 0.26$ & 9 & 
	$-18.44 \pm 0.26$ &
	$8.21 \pm 0.05$ & 12 & 
	2.62 & 8.13 
	\\
NGC 2366 &
        G69 & \nodata & 
	$35.2 \pm 3.5$ &
	$27.52 \pm 0.28$ & 1,10 & 
	$-18.18 \pm 0.28$ &
	$7.91 \pm 0.08$ & 13,14 & 
	2.18 & 7.91 
	\\
NGC 3109 &
        G128 & 5052928 & 
	$131 \pm 13$ &
	$25.52 \pm 0.18$ & 11--13 & 
	$-17.61 \pm 0.10$ & 
	$8.06 \pm 0.20$ & 7,15,16 & 
	1.11 & 7.41 
	\\
NGC 3738 &
        G69 & \nodata & 
	$32.8 \pm 3.3$ &
	$28.45 \pm 0.25$ & 14 & 	
	$-19.03 \pm 0.25$ &
	$8.23 \pm 0.01$ & 11 & 
	2.45 & 8.32 
	\\
NGC 4214 &
        G69 & \nodata & 
	$175 \pm 18$ &
	$27.13 \pm 0.20$ & 15 & 
	$-19.53 \pm 0.20$ &
	$8.22 \pm 0.05$ & 17 & 
	2.55 & 8.55 
	\\
NGC 4449 &
        G69 & \nodata & 
	$303 \pm 30$ &
	$28.12 \pm 0.27$ & 14 & 	
	$-21.12 \pm 0.27$ &
	$8.31 \pm 0.07$ & 11,18 & 
	2.98 & 9.29 
	\\
NGC 5408 &
        S & \nodata & 
	$27.2 \pm 2.7$ & 
	$28.41 \pm 0.17$ & 16 & 
	$-18.79 \pm 0.17$ &
	$7.98 \pm 0.01$ & 19,20 & 
	2.29 & 8.19 
	\\
NGC 55 &
        G128 & 5056000 & 
	$785 \pm 79$ &		
	$25.85 \pm 0.20$ & 17--19 & 
	$-19.88 \pm 0.20$ &	
	$8.05 \pm 0.10$ & 21--23 & 
	1.58 & 8.44 
	\\
NGC 6822 &
        S & \nodata & 
	$1440 \pm 140$ &
        $23.49 \pm 0.08$ & 20--22 & 
	$-18.14 \pm 0.10$ &
	$8.11 \pm 0.10$ & 24,25 &
	2.98 & 8.10 
	\\
Peg DIG$\,$\tablenotemark{a} & 
        G128 & 5055744 & 
	$30.5\pm 3.1$ &
	$24.90 \pm 0.10$ & 23 & 
	$-15.40 \pm 0.10$ & 
	$7.93 \pm 0.13$ & 26 & 
	2.90 & 6.98 
	\\ 
Sextans A &
        G128 & 5053696 & 
	$8.39 \pm 0.84$ &
	$25.75 \pm 0.15$ & 24--26 & 
	$-14.85 \pm 0.15$ & 
	$7.54 \pm 0.06$ & 1,27,28 & 
	0.85 & 6.24 
	\\
Sextans B &
        G128 & 5052672 & 
	$23.0 \pm 2.3$ &
	$25.63 \pm 0.04$ & 6,13,24,27 &
	$-15.83 \pm 0.10$ & 
	$7.53 \pm 0.05$ & 15,27--29 & 
	1.73 & 6.86 
	\\
UGC 6456 &
        G59 & \nodata & 
	$2.25 \pm 0.23$ &
	$28.19 \pm 0.04$ & 13,28,29 & 	
	$-15.87 \pm 0.10$ &
	$7.69 \pm 0.07$ & 11,14,30,31 &
	1.83 & 6.90 
	\\
WLM &
        G128 & 5051136 & 
	$45.8 \pm 4.6$ &
	$24.83 \pm 0.08$ & 31--33 & 
	$-15.78 \pm 0.10$ &
	$7.83 \pm 0.06$ & 32 & 
	1.88 & 6.88 
\\[1mm]
\tableline
\vspace*{5mm}
\end{tabular}
\tablenotetext{a}{
Alternative names: 
GR~8 --- DDO~155, UGC~8091; 
M81 dwB --- UGC~5423;
Pegasus dwarf irregular (Peg DIG) --- DDO~216, UGC~12613.
}
\tablenotetext{b}{
Ho~IX: 
We have adopted the distance to M~81 \citep{ferrarese00}.
\cite{miller95} measured 12$+$log(O/H) $\sim 8.0 \pm 0.2$ 
from a supernova remnant.
}
\tablecomments{
Col.~(1): Galaxy name (alphabetical order). 
Col.~(2): Data source: 
G59 --- GTO~59 (P.I. G.~Rieke),
G69 --- GTO~69 (P.I. G.~Fazio), 
G128 --- GTO~128 (P.I. R.~D.~Gehrz), and 
S --- SINGS 
(159; \citealp{sings}).
Col.~(3): AOR Key pertaining to galaxies observed in program GTO~128.
Col.~(4): Flux density at 4.5 \micron.
Our measured flux densities for SINGS galaxies are consistent with
those found in \cite{dale05}; see also Fig.~\ref{fig_dale}. 
Cols.~(5) and (6): Distance modulus and references, respectively.
Col.~(7): Absolute magnitude at 4.5~\micron.
Cols.~(8) and (9): Nebular (H~II region) oxygen abundance and
references, respectively. 
Col.~(10): Derived \btwo\ colors in the present work, where $B$
magnitudes are corrected for Galactic extinctions; see \cite{kara04}.
Col.~(11): Derived stellar mass using model~8 in
Table~\ref{table_masses}, and adjusted downwards by 0.25~dex for the
\cite{chabrier03} IMF.
}
\vspace*{3mm}
\tablerefs{
Distance references:
1.~\citealt{makarova98}; 
2.~\citealt{kara02a}; 
3.~\citealt{dp98}; 
4.~\citealt{cole99}; 
5.~\citealt{dolphin01}; 
6.~\citealt{kara02c}; 
7.~\citealt{tolstoy98}; 
8.~\citealt{angeretti05}; 
9.~\citealt{tosi01}; 
10.~\citealt{ti05}; 
11.~\citealt{musella97}; 
12.~\citealt{minniti99}; 
13.~\citealt{mendez02}; 
14.~\citealt{kara03a}; 
15.~\citealt{drozd02}; 
16.~\citealt{kara02b}; 
17.~\citealt{pritchet87}; 
18.~\citealt{kara03b}; 
19.~\citealt{tgd05}; 
20.~\citealt{gallart96}; 
21.~\citealt{clementini03}; 
22.~\citealt{pietr04}; 
23.~\citealt{aparicio94}; 
24.~\citealt{piotto94}; 
25.~\citealt{sakai96}; 
26.~\citealt{dolphin03}; 
27.~\citealt{sakai97}; 
28.~\citealt{lynds98}; 
29.~\citealt{slmh98}; 
30.~\citealt{mz97}; 
31.~\citealt{dolphin00}; 
32.~\citealt{rejkuba00}. 
\newline\newline
Spectroscopic oxygen abundance references:
1.~\citealt{vanzee97}; 
2.~\citealt{ks01}; 
3.~\citealt{skh89}; 
4.~\citealt{vanzee06}; 
5.~\citealt{mh96}; 
6.~\citealt{mmdo91}; 
7.~\citealt{lee03field}; 
8.~\citealt{lee03south}; 
9.~\citealt{drd97}; 
10.~\citealt{ks97}; 
11.~\citealt{martin97}; 
12.~\citealt{ls04}; 
13.~\citealt{gd94}; 
14.~\citealt{itl97}; 
15.~\citealt{lzg06}; 
16.~\citealt{rm95}; 
17.~\citealt{ks96}; 
18.~\citealt{boeker01}; 
19.~\citealt{scv86}; 
20.~\citealt{mmca94}; 
21.~\citealt{talent80}; 
22.~\citealt{ws83}; 
23.~\citealt{tuellmann03}; 
24.~\citealt{ppr05}; 
25.~\citealt{lsv06}; 
26.~\citealt{sbk97}; 
27.~\citealt{kniazev05}; 
28.~\citealt{magrini05}; 
29.~\citealt{moles90}; 
30.~\citealt{lynds98}; 
31.~\citealt{hh99}; 
32.~\citealt{lsv05}. 
} 
\end{center}
\end{table}


\begin{table}
\begin{center}
\setlength{\tabcolsep}{2.mm}
\renewcommand{\arraystretch}{1.3}
\caption{
Comparison among selected luminosity-metallicity relations for
nearby dwarf irregular galaxies and other distant star-forming
galaxies. 
We show here a number of \lz\ relations from the literature;
this is not intended to be a complete listing.
We have not performed a homogeneous treatment of the data listed in
this table.
\vspace*{3mm}
\label{table_lzfits}
}
\begin{tabular}{lccccccc}
\tableline \tableline
& No. of & Luminosity & Wavelength & 
\multicolumn{2}{c}{\lz\ fits} & & Dispersion in \\
Data source & Galaxies & Range &
(\micron) & $a$ & $b$ & $a_{16}$ & log(O/H) (dex) \\
(1) & (2) & (3) & (4) & (5) & (6) & (7) & (8)
\\[1mm]
\tableline
\multicolumn{8}{c}{{\sf Local galaxies ($D \la 10$~Mpc)}} \\
\tableline
This paper & 25 & 
	$-11 \ga M_B \ga -18$ & 4.5 \phm{($X$)} & 
	$5.78 \pm 0.21$ & $-0.122 \pm 0.012$ & 7.73 &
	0.12 \\ 
This paper & 25 & 
	$-11 \ga M_B \ga -18$ & 0.4 ($B$) & 
	$5.94 \pm 0.27$ & $-0.128 \pm 0.017$ & 7.99 &
	0.16 \\ 
\citealt{oliveira06} & 29 & 
	$-12 \ga M_B \ga -18$ & \phn2.2 ($K_s$) &
	$5.55 \pm 0.26$ & $-0.14 \pm 0.02$ & 7.79 &
	0.15 \\ 
\citealt{vanzee06} & 33 & 
	$-11 \ga M_B \ga -18$ & 0.4 ($B$) &
	$5.65 \pm 0.17$ & $-0.149 \pm 0.011$ & 8.03 &
	0.15 \\ 
\citealt{lee03field} & 22 & 
	$-11 \ga M_B \ga -18$ & 0.4 ($B$) & 
	$5.59 \pm 0.54$ & $-0.153 \pm 0.025$ & 8.04 &
	0.18 \\ 
\citealt{rm95} & 21 & 
	$-11 \ga M_B \ga -18$ & 0.4 ($B$) &
	$5.67 \pm 0.48$ & $-0.147 \pm 0.029$ & 8.02 &
	0.08$\,$\tablenotemark{a} \\ 
\citealt{skh89} & 20 & 
	$-11 \ga M_B \ga -19$ & 0.4 ($B$) &
	5.50 & $-0.153$ & 7.95 &
	0.16 \\[1mm] 
\tableline
\multicolumn{8}{c}{{\sf Distant galaxies}} \\
\tableline
\citealt{rosenberg06} & 17 & 
	$-16 \ga M_B \ga -18$ & 3.6 \phm{($B$)} & 
	4.85 & $-0.16$ & 7.41 &
	0.16 \\ 
\citealt{salzer05} & 370 & 
	$-12 \ga M_B \ga -22$ & 2.2 ($K$) &
	$3.92 \pm 0.09$ & $-0.212 \pm 0.003$ & 7.31 &
	0.24 \\ 
\citealt{savaglio05} & 79 & 
        $-18 \ga M_B \ga -22$ & 0.4 ($B$) &
        $2.98 \pm 0.94$ & $-0.280 \pm 0.045$ & 7.46 &
	\nodata \\
\citealt{kk04} & 177 & 
        $-17 \ga M_B \ga -22$ & 0.4 ($B$) &
        4.90 & $-0.193$ & 7.99 &
	\nodata \\
\citealt{jlee04} & 54 & 
	$-12 \ga M_B \ga -20$ & 0.4 ($B$) &
	$5.37 \pm 0.46$ & $-0.159 \pm 0.029$ & 7.91 &
	0.26 \\ 
\citealt{tremonti04} & $\ga 53000$ & 
	$-16 \ga M_B \ga -22$ & \phd0.4 ($B$)$\,$\tablenotemark{b} &
	$5.238 \pm 0.018$ & $-0.185 \pm 0.001$ & 8.20 &
	0.16 \\
%
\tableline
\end{tabular}
\vspace*{3mm}
\tablenotetext{a}{
\cite{rm95} stated the fit parameters and resulting dispersion were 
effective only for galaxies brighter than $M_B = -15$, as the
dispersion increased towards lower luminosities. 
However, \cite{lee03field} showed that improved data
(e.g., distances) reduced the dispersion at lower luminosities.
}
\tablenotetext{b}{
\cite{tremonti04} obtained galaxy luminosities in Sloan filter $g$
(similar to $B$), and accounted for the appropriate transformations
from $g$ to $B$.
}
\tablecomments{
Col.~(1): Reference.
Col.~(2): Number of galaxies.
Col.~(3): Luminosity range.  
A number of the studies quote $B,AB$ magnitudes, which
are $0.163$ mag brighter than Johnson $B$ magnitudes 
\citep{fg94}.
Col.~(4): Wavelength.
Cols.~(5) and (6): Fits to the respective relations are expressed as
12$+$log(O/H) = $a + b\,M_{\lambda}$, where $a$ and $b$ are the
intercept and the slope, respectively, and $M_{\lambda}$ is the
absolute magnitude at wavelength $\lambda$.
Col.~(7): $a_{16}$ represents the oxygen abundance determined from the
specified relation at a fiducial galaxy luminosity 
equal to $M_{\lambda} = -16$.
Col.~(8): Dispersion in oxygen abundance.
}
%
\end{center}
\end{table}


\begin{table}
\begin{center}
\setlength{\tabcolsep}{2.mm}
\renewcommand{\arraystretch}{1.3}
\caption{
Stellar masses using models of stellar mass-to-light
ratios from Tables 3 and 4 in \cite{belldejong01}
and the method described in \S~\ref{sec_mz}.
From the resulting derived stellar masses, we performed
the resulting \mz\ fit for each model.
We list each \mz\ fit as 12$+$log(O/H) = 
$a + b$ log \mstar, and we list the dispersion in
oxygen abundance for each fit.
\vspace*{3mm}
\label{table_masses}
}
\begin{tabular}{lcccccccccc}
\tableline \tableline
& & \multicolumn{2}{c}{\bv} & \multicolumn{2}{c}{\vk} &
\multicolumn{2}{c}{\bk} & \multicolumn{2}{c}{\mz\ fits} &
Dispersion in \\
Model \# & Model & 
$a_K$ & $b_K$ & $a_K$ & $b_K$ & $a^{\prime}$ & $b^{\prime}$ & 
$a$ & $b$ & log(O/H) (dex) \\
(1) & (2) & (3) & (4) & (5) & (6) & (7) & (8) & (9) & (10) & (11)
\\[1mm]
\tableline
\multicolumn{11}{c}{\sf Scaled Salpeter IMF, \zsun} \\
\tableline
1 & Closed box & 
    $-$0.554 & 0.540 & $-$0.740 & 0.207 & $-$0.688 & 0.150 &
    $5.52 \pm 0.24$ & $0.307 \pm 0.031$ & 0.118
\\
2 & Infall & 
    $-$0.692 & 0.699 & $-$0.926 & 0.258 & $-$0.863 & 0.188 &
    $5.57 \pm 0.23$ & $0.304 \pm 0.030$ & 0.117
\\
3 & Outflow & 
    $-$0.534 & 0.500 & $-$0.622 & 0.157 & $-$0.601 & 0.119 &
    $5.49 \pm 0.24$ & $0.309 \pm 0.031$ & 0.118
\\
4 & Dynamical time & 
    $-$0.531 & 0.476 & $-$0.687 & 0.177 & $-$0.645 & 0.129 &
    $5.50 \pm 0.23$ & $0.309 \pm 0.030$ & 0.118
\\
5 & Formation epoch & 
    $-$0.694 & 0.676 & $-$0.966 & 0.270 & $-$0.888 & 0.193 &
    $5.58 \pm 0.23$ & $0.303 \pm 0.030$ & 0.117
\\
6 & Form. epoch: bursts & 
    $-$0.692 & 0.652 & $-$1.087 & 0.314 & $-$0.959 & 0.212 &
    $5.60 \pm 0.23$ & $0.302 \pm 0.030$ & 0.117
\\
7 & Cole et al. (2000) & 
    $-$0.654 & 0.696 & $-$1.019 & 0.301 & $-$0.909 & 0.210 &
    $5.58 \pm 0.23$ & $0.302 \pm 0.030$ & 0.117
\\[1mm]
\tableline
\multicolumn{11}{c}
{\sf Different SPS$\,$\tablenotemark{a} Models (IMFs); 0.4 \zsun} \\
\tableline
8 & BC Salpeter & 
    $-$0.43 & 0.60 & $-$1.16 & 0.44 & $-$0.851 & 0.254 &
    $5.58 \pm 0.23$ & $0.298 \pm 0.030$ & 0.117
\\

9 & BC scaled Salpeter & 
    $-$0.59 & 0.60 & $-$1.31 & 0.44 & $-$1.005 & 0.254 &
    $5.62 \pm 0.23$ & $0.298 \pm 0.030$ & 0.117
\\
10 & BC modif. Salpeter & 
     $-$0.71 & 0.60 & $-$1.45 & 0.45 & $-$1.133 & 0.257 &
     $5.66 \pm 0.22$ & $0.297 \pm 0.029$ & 0.117
\\
11 & BC96 Scalo & 
     $-$0.65 & 0.56 & $-$1.36 & 0.43 & $-$1.052 & 0.243 &
     $5.64 \pm 0.23$ & $0.299 \pm 0.030$ & 0.117
\\
12 & KA Salpeter & 
     $-$0.38 & 0.53 & $-$1.03 & 0.39 & $-$0.754 & 0.225 &
     $5.55 \pm 0.24$ & $0.300 \pm 0.030$ & 0.117
\\
13 & Schulz et al. Salpeter & 
     $-$0.65 & 0.77 & $-$2.50 & 0.81 & $-$1.552 & 0.395 &
     $5.81 \pm 0.22$ & $0.284 \pm 0.030$ & 0.120
\\
14 & P\'egase Salpeter & 
     $-$0.38 & 0.59 & $-$1.15 & 0.45 & $-$0.817 & 0.255 &
     $5.57 \pm 0.24$ & $0.298 \pm 0.030$ & 0.117
\\
15 & P\'egase $x=-1.85$ & 
     $-$0.07 & 0.48 & $-$0.71 & 0.37 & $-$0.431 & 0.209 &
     $5.45 \pm 0.24$ & $0.301 \pm 0.030$ & 0.117
\\
16 & P\'egase $x=-0.85$ & 
     $-$0.59 & 0.71 & $-$1.25 & 0.45 & $-$0.994 & 0.275 &
     $5.62 \pm 0.23$ & $0.296 \pm 0.030$ & 0.117
\\[1mm]
\tableline
\end{tabular}
\vspace*{3mm}
\tablenotetext{a}{
SPS models - stellar population synthesis models.
}
\tablecomments{
Col.~(1): Model number.
Col.~(2): Model type. 
BC -- Bruzual \& Charlot, KA -- Kodama \& Arimoto;
see \cite{belldejong01} for details.
Cols.~(3) and (4): Intercept and slope, respectively, for 
log~(\mstar/$L_K$) as a linear function of \bv.
Cols.~(5) and (6): Intercept and slope, respectively, for 
log~(\mstar/$L_K$) as a linear function of \vk.
Cols.~(7) and (8): Derived intercept and slope, respectively, for 
log~(\mstar/$L_K$) as a linear function of \bk.
Cols.~(9) and (10): Derived intercept and slope, respectively, for the
subsequent \mz\ fits for the given model.
Cols.~(11): Dispersion in log(O/H). 
}
%
\end{center}
\end{table}


\begin{table}
\begin{center}
\setlength{\tabcolsep}{2mm}
\renewcommand{\arraystretch}{1.3}
\caption{
Comparison among selected stellar mass-metallicity relations for
nearby dwarf irregular galaxies and star-forming galaxies
at distant epochs.
Fits to the respective relations are expressed as
12$+$log(O/H) = $f(X)$, or functions of $X$ = log \mstar.
We have not homogenized measures of the stellar mass or
metallicity for the other studies.
This is not a complete listing, as we have selected recently published
\mz\ relations from the literature.
\vspace*{5mm}
\label{table_mzfits}
}
\begin{tabular}{lccccc}
\tableline \tableline
& No. of & & Luminosity & \mz\ fits & Dispersion in \\
Data source & Galaxies & Redshift & Range & ($X$ = log \mstar) & 
log(O/H) (dex) 
\\[1mm]
\tableline
This paper & 25 & $\la 0.001$ &
	$-13 \ga M_{[4.5]} \ga -21$ &  
	$(5.54 \pm 0.24) + (0.298 \pm 0.030)\,X$ & 0.12 \\
\citealt{savaglio05} & 56 & $\sim 0.7$ &
	$-18 \ga M_{B,AB} \ga -22$ & 
	$(4.06 \pm 0.58) + (0.478 \pm 0.058)\,X$ & 0.20 \\
\citealt{tremonti04} & $\ga 53000$ & $\sim 0.1$ &
	$-16 \ga M_B \ga -22$ & 
	$-1.492 + 1.847 X - 0.080 X^2$ & 0.10 \\
\citealt{pg03} & 163 & $\sim 0.026$ &
        $-19 \ga M_K \ga -26$ &
	$(6.02 \pm 0.28) + (0.257 \pm 0.028)\,X$ & 0.21 \\[1mm]
\tableline
\end{tabular}
\vspace*{-5mm}
%
%
\end{center}
\end{table}


\begin{table}
\begin{center}
\setlength{\tabcolsep}{2mm}
\renewcommand{\arraystretch}{1.2}
\caption{
Gas-to-stellar mass fractions and effective yields.
\vspace*{3mm}
\label{table_gas}
}
\begin{tabular}{lcccc}
\tableline \tableline
& $F_{21}$ & & & $v_{\rm rot}$ \\
Galaxy & (Jy km s$^{-1}$) & log ($M_{\rm gas}$/\mstar) & 
log $y_{\rm eff}$ & (km s$^{-1}$) \\
(1) & (2) & (3) & (4) & (5)
\\[1mm]
\tableline
DDO 154 & 
        145 & 2.00 & $-$1.26 & 54
	\\
DDO 165 & 
        28.2 & 0.55 & \nodata & 23
	\\
DDO 53 & 
        13.8 & 0.770 & $-$2.51 & 17
	\\
GR 8 & 
        8.6 & 0.464 & $-$2.75 & 21
	\\
Ho I & 
        38.9 & 1.18 & $-$2.03 & 18
	\\
Ho II & 
        363 & 1.17 & $-$1.82 & 56
	\\
Ho IX & 
        158 & 1.93 & \nodata & 50
	\\
IC 1613 & 
        698 & 1.26 & $-$2.04 & 19
	\\
IC 2574 & 
        447 & 0.97 & $-$1.90 & 55
	\\
IC 5152 & 
        98.0 & 0.02 & $-$2.84 & 50
	\\
Leo A & 
        68.4 & 1.13 & $-$2.49 & 4
	\\
M81 dwB & 
        3.8 & 0.34 & $-$2.52 & 25
	\\
NGC 1569 & 
        74.1 & $-$0.01 & $-$2.59 & 37
	\\
NGC 1705 & 
        16.6 & 0.02 & $-$2.55 & 61
	\\
NGC 2366 & 
        295 & 1.07 & $-$1.93 & 45
	\\
NGC 3109 & 
        1110 & 1.39 & $-$1.47 & 51
	\\
NGC 3738 & 
        21.9 & $-$0.09 & $-$2.61 & 50
	\\
NGC 4214 & 
        324 & 0.32 & $-$2.30 & 43
	\\
NGC 4449 & 
        794 & 0.36 & $-$2.18 & 84
	\\
NGC 5408 & 
        64.6 & 0.49 & $-$2.37 & 33
	\\
NGC 55 & 
        2679 & 1.15 & $-$1.72 & 78
	\\
NGC 6822 & 
        2399 & 0.17 & $-$2.54 & 55
	\\
Peg DIG & 
        22.8 & $-$0.36 & $-$3.07 & 8
	\\ 
Sextans A & 
        208.8 & 1.83 & $-$1.56 & 49
	\\
Sextans B & 
        102.4 & 0.92 & $-$2.45 & 22
	\\
UGC 6456 & 
        14.1 & 1.03 & $-$2.19 & 7
	\\
WLM & 
        299.8 & 1.03 & $-$2.05 & 26
\\[1mm]
\tableline
\vspace*{5mm}
\end{tabular}
\tablecomments{
Col.~(1): Galaxy name.
Col.~(2): 21-cm flux densities \citep{kara04}.
Col.~(3): Gas-to-stellar mass ratio.
Col.~(4): Logarithm of the effective yield.
Col.~(5): H~I rotational velocity corrected for inclination and
turbulent motions \citep{kara04}.
}
\end{center}
\end{table}

\clearpage	

\begin{figure}
\caption{
Postage-stamp images at 4.5 \micron\ of dwarf galaxies
from GTO~128.
North and East are at the top and left, respectively, in each frame.
All galaxies are displayed with the same isophotal scaling to
emphasize the large range in surface brightness of the dwarf galaxies.
The fields of view shown are:
GR~8 --- $5\arcmin\,\times\,5\arcmin$,
IC~1613 --- $10\arcmin\,\times\,10\arcmin$,
IC~5152 --- $5\arcmin\,\times\,5\arcmin$,
Leo~A --- $5\arcmin\,\times\,5\arcmin$,
NGC~55 --- $15\arcmin\,\times\,10\arcmin$,
NGC~3109 --- $10\arcmin\,\times\,5\arcmin$,
Peg~DIG --- $5\arcmin\,\times\,5\arcmin$,
Sextans~A --- $5\arcmin\,\times\,5\arcmin$,
Sextans~B --- $5\arcmin\,\times\,5\arcmin$, and
WLM --- $5\arcmin\,\times\,15\arcmin$.
}
\label{fig_stamps}
\end{figure}


\begin{figure}
\caption{
Example of galaxy photometry for Sextans~B.
North and East are to the top and left, respectively,
in the 4.5~\micron\ image.
The central elliptical aperture surrounds the galaxy, and
the seven circular apertures sample the background flux.
The galaxy aperture is aligned roughly to the same orientation
as the optical emission seen on images from the Digitized Sky Survey. 
}
\label{fig_gxyphot}
\end{figure}


\begin{figure}
\epsscale{0.72} 
\plotone{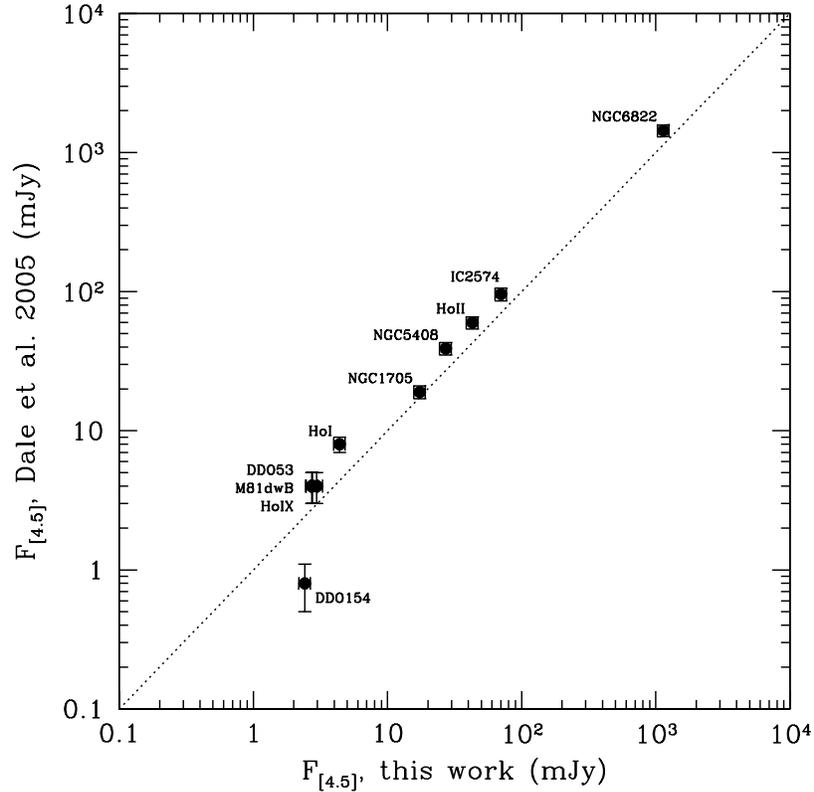} 
\caption{
Comparison of total 4.5~\micron\ fluxes for selected
dwarf galaxies from the SINGS sample: fluxes from \cite{dale05} versus
measured fluxes from the present work. 
The dotted line marks the line of equal fluxes.
}
\label{fig_dale}
\end{figure}


\begin{figure}
\epsscale{0.72} 
\plotone{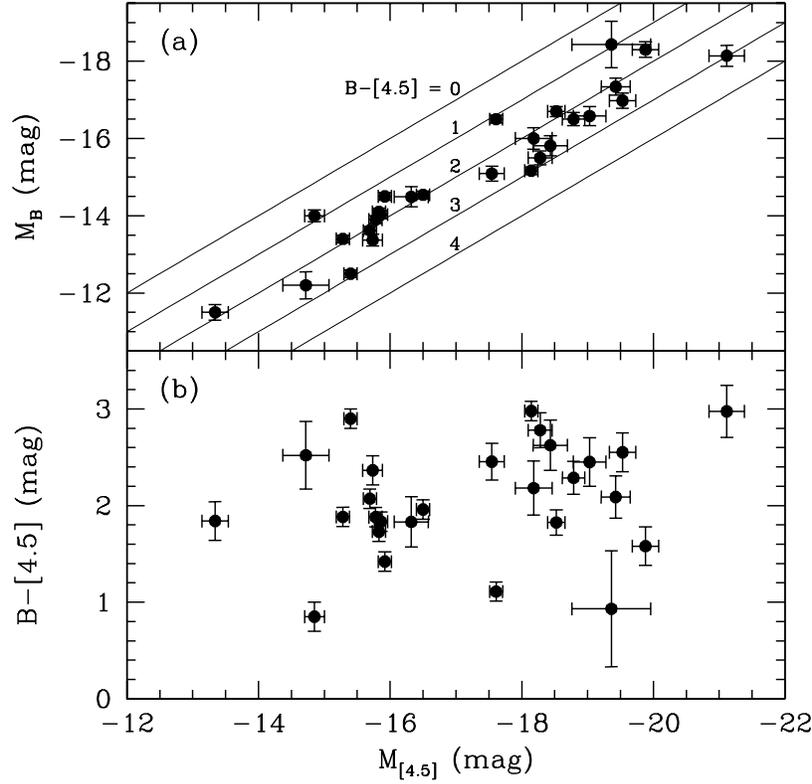} 
\caption{
(a) $B$ absolute magnitude versus [4.5] absolute magnitude.
Lines of constant \btwo\ color between 0 and 4 are also plotted.
(b) \btwo\ color versus [4.5] absolute magnitude.
The Pearson correlation coefficient for this color-magnitude
relation is $r = -0.24$.
}
\label{fig_lcomp}
\end{figure}


\begin{figure}
\epsscale{0.72} 
\plotone{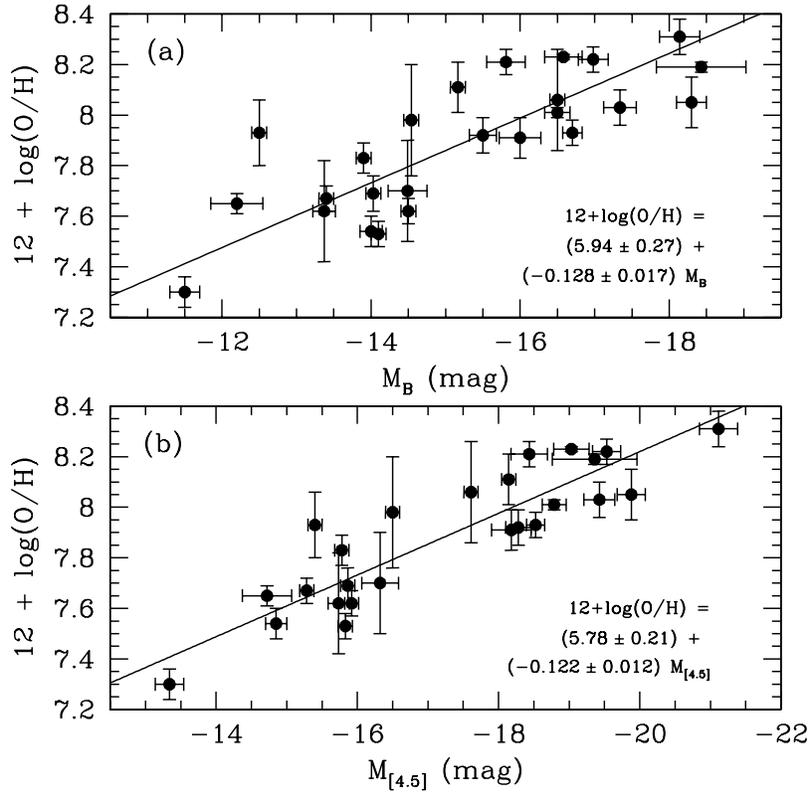} 
\caption{
Nebular oxygen abundances versus 
(a) absolute magnitude in $B$, and
(b) absolute magnitude at 4.5~\micron.
The solid lines represent fits to the data, whose fit parameters are
given in each panel.
The Pearson coefficients are $r = -0.80$ in $B$, and 
$r = -0.89$ in [4.5].
}
\label{fig_oxyll}
\end{figure}


\begin{figure}
\epsscale{0.70} 
\plotone{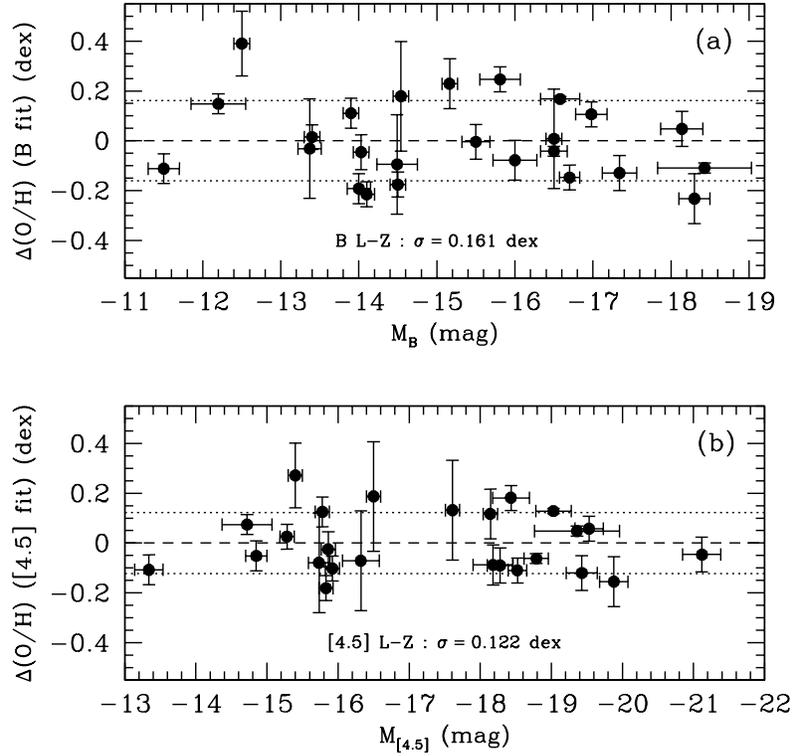} 
\caption{
(a) Residuals in oxygen abundance (from $B$ fit) versus
$B$ absolute magnitude.
(b) Residuals in oxygen abundance (from [4.5] fit) versus
[4.5] absolute magnitude.
Both panels are plotted to the same vertical scale, and
a short-dashed horizontal line in each panel indicates zero residual.
The dotted horizontal lines in each panel indicate the $1\sigma$
scatter.
The scatter in the NIR \lz\ relation is reduced compared to that found
in the $B$ \lz\ relation.
}
\label{fig_offmag}
\end{figure}


\begin{figure}
\epsscale{0.68} 
\plotone{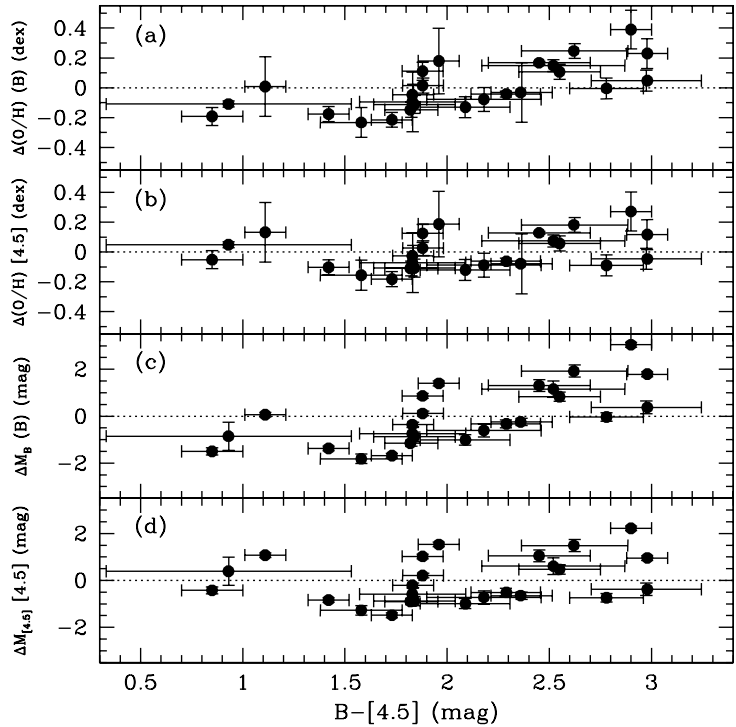} 
\caption{
Panel (a): Residuals in oxygen abundance from the $B$ \lz\ fit versus
\btwo\ color.
Panel (b): Residuals in oxygen abundance from the [4.5] \lz\ fit
versus \btwo\ color.
Panels (a) and (b) are plotted to the same vertical scale.
Panel (c): Residuals in $B$ absolute magnitude from the $B$ \lz\ fit
versus \btwo\ color.
Panel (d): Residuals in [4.5] absolute magnitude from the [4.5] \lz\
fit versus \btwo\ color.
Panels (c) and (d) are plotted to the same vertical scale.
Residuals at optical wavelengths correlate ($r = +0.67$) more strongly
with \btwo\ color than residuals at NIR wavelengths ($r = +0.28$).
}
\label{fig_offcol}
\end{figure}


\begin{figure}
\epsscale{0.68} 
\plotone{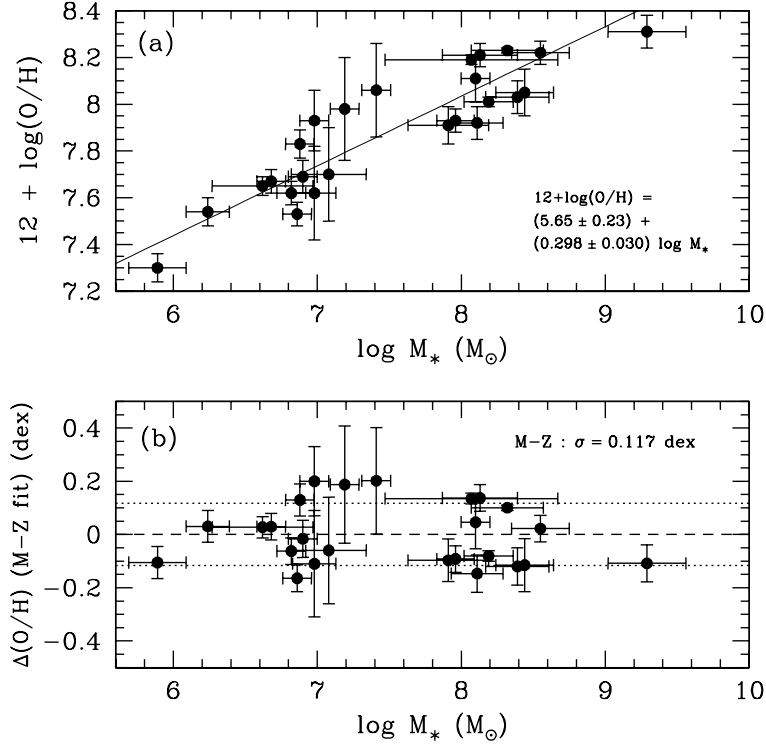} 
\caption{
Panel (a): oxygen abundance versus stellar mass.
Stellar masses have been derived using the color-based stellar
mass-to-light ratios from model 8 in Table~\ref{table_masses}, and
have been corrected to the \cite{chabrier03} IMF.
The fit to the data is shown as the solid line;
the Pearson coefficient is $r = +0.90$.
Panel (b): residuals in oxygen abundance from our present
\mz\ relation against stellar mass.
The Pearson coefficient is $r = -0.08$.
}
\label{fig_zmstar}
\end{figure}


\begin{figure}
\epsscale{0.65} 
\plotone{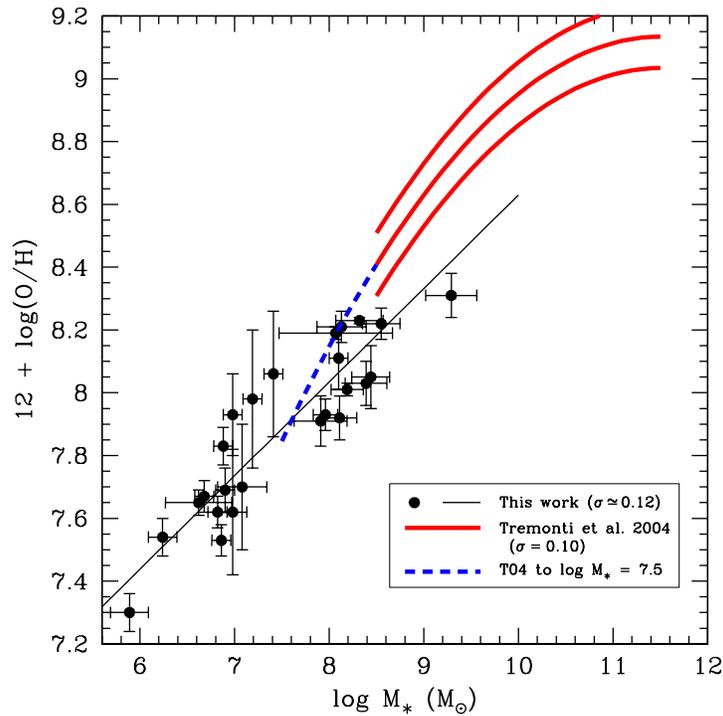} 
\caption{
Oxygen abundance versus stellar mass.
The plot has been expanded to include more massive galaxies.
Data and the least-squares fit to the present data from
Fig.~\ref{fig_zmstar} are shown; the fit is shown as a thin black
solid line and is arbitrarily extrapolated to $10^{10}\,$\msun.
The polynomial fit to over 53000 galaxies from the SDSS
\citep{tremonti04} is shown as a heavy red solid line; 
the $1\sigma$ curves are also shown.
Their fit is effective over the mass range between
8.5 $\la$ log~(\mstar/\msun) $\la$ 11.5.
In the ``overlap region'' 
(8.5 $\la$ log~(\mstar/\msun) $\la$ 9.5), SDSS galaxies are
more metal-rich by $\approx$ 0.2--0.3~dex compared to dwarf
galaxies in the present sample.
An extrapolation of the SDSS \mz\ relation down to 
log~(\mstar/\msun) = 7.5 is shown as a blue short-dashed line.
}
\label{fig_sdss}
\end{figure}


\begin{figure}
\caption{
Panels (a) and (b): oxygen abundances are plotted against
gas-to-stellar mass ratio.
In panel (a), the solid line represents the fit to the data;
the Pearson coefficient is $r = -0.58$.
In panel (b), we have plotted outflow variations of the closed-box
model, where the outflow rate is a constant multiple of the
star-formation rate.
The purple and blue curves describe models with
$y_O = 0.01$ \citep{tremonti04} and 
$y_O = 2 \times 10^{-3}$ (e.g., \citealp{lee03field,vzh06}),
respectively.
For each set of curves, the outflow rates from right to left are 
0, 5, 10, 25, and 50 times the star-formation rate, respectively.
In panel~(c), \btwo\ colors are plotted against 
gas-to-stellar mass ratio.
The solid line represents the fit, and the Pearson coefficient is 
$r = -0.83$.
The fit excludes DDO~165, Ho~IX, and NGC~1569; the latter
is the outlier with a very blue color for its relatively 
low gas-to-stellar mass ratio.
}
\label{fig_gasstar}
\end{figure}


\begin{figure}
\epsscale{0.62} 
\plotone{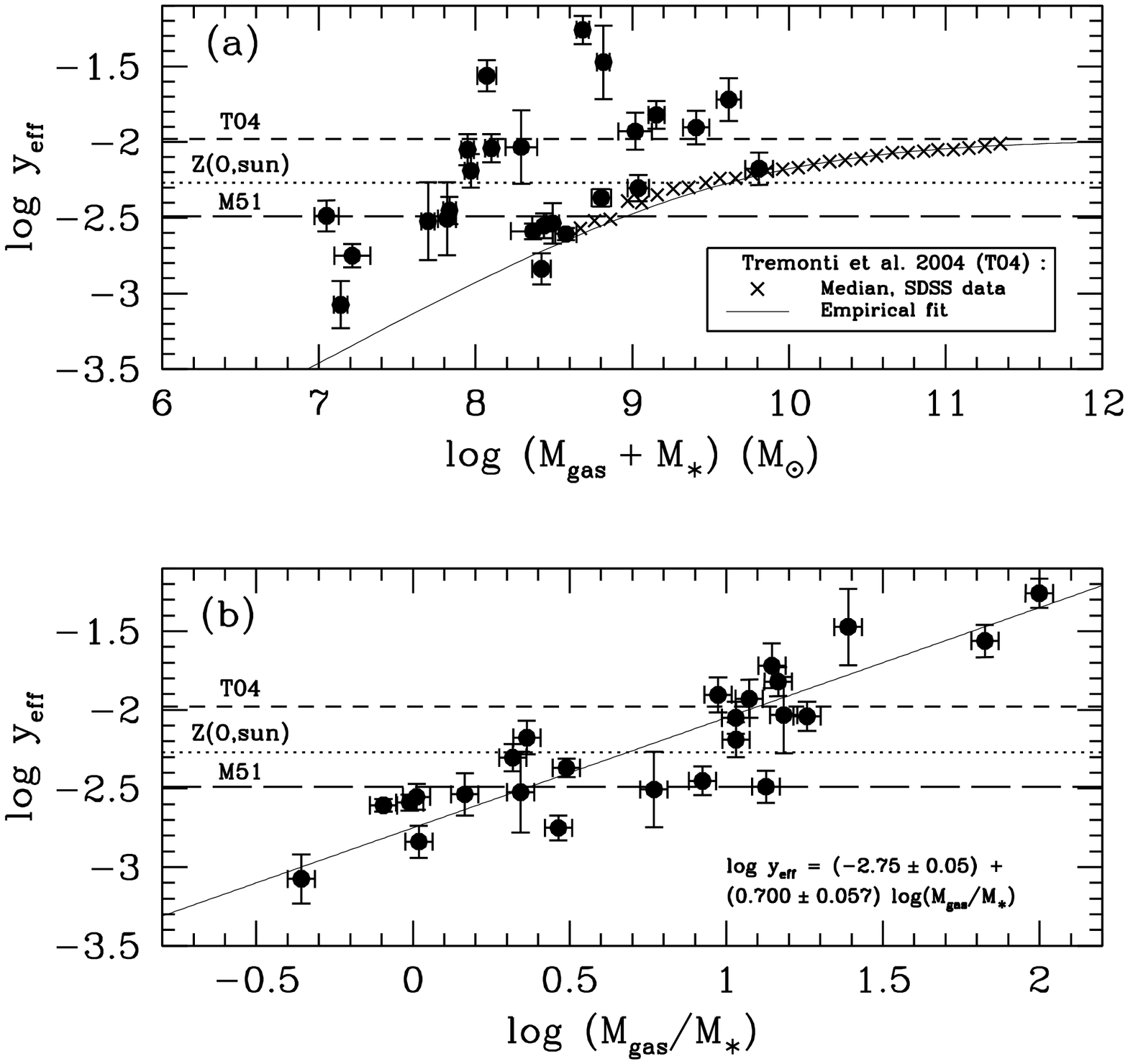} 
\caption{\small
Panel (a): effective yield against total baryonic mass in gas and
stars; cf. also \cite{garnett02}. 
Filled circles represent dwarf galaxies in the present work.
Crosses represent the median of the SDSS data in mass bins of
0.1~dex \citep{tremonti04}.
The empirical fit is shown as a thin solid line 
\citep[][their Equation 6]{tremonti04}.
The asymptotic yield ($y_{\rm eff} = 0.0104$)
is represented by the short-dashed line.
The solar oxygen mass fraction \citep{asplund04,melendez04}
and the effective yield for the nearby spiral galaxy M51
\citep{bgk04} are represented by
dotted and long-dashed lines, respectively.
These values are comparable to the range of effective yields for dwarf
galaxies (e.g., log $y_{\rm eff} \simeq$ $-2.7$ to $-2.1$, 
\citealp{lee03field,vzh06}).
Panel~(b): effective yield versus gas-to-stellar mass ratio.
The fit shown is represented by a thin solid line.
The trend here is similar to the one observed for a sample
of nearby UGC dwarf galaxies described by \cite{vzh06}.
The horizontal lines are the same as in the panel above.
}
\label{fig_yield}
\end{figure}


\begin{figure}
\epsscale{0.62} 
\plotone{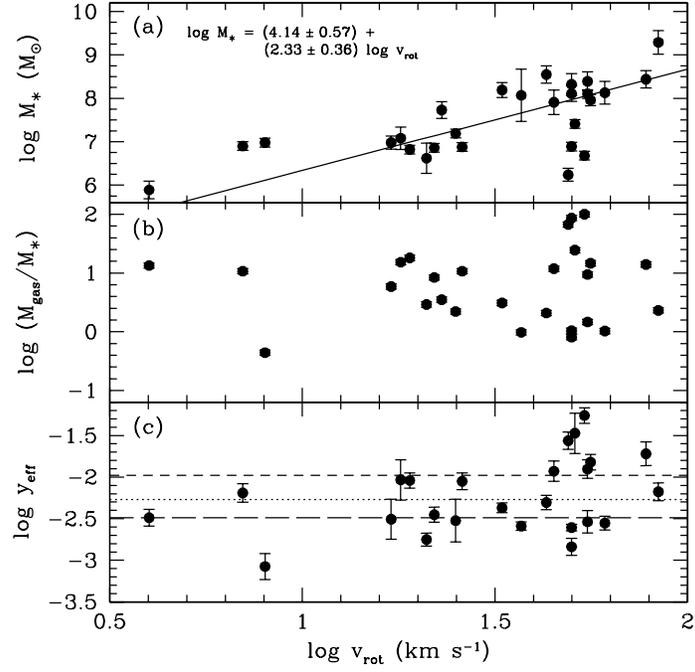} 
\caption{
Panel (a): stellar mass versus \hi\ rotation velocity.
The solid line represents the fit to the data
with Pearson coefficient $r = +0.67$.
Panel (b): gas-to-stellar mass ratio versus \hi\ rotation velocity.
Panel (c): effective yield versus \hi\ rotation velocity.
\hi\ rotation velocities in Table~\ref{table_sample}
have been corrected for inclination and turbulent 
motions; see \cite{kara04}.
The Pearson correlation coefficient between
rotation velocity and effective yield is $r = +0.40$.
The horizontal lines are the same as in Fig.~\ref{fig_yield}.
}
\label{fig_vrot}
\end{figure}

\end{document}